\documentclass[aps,pre,superscriptaddress,amssymb]{revtex4-1}
\usepackage{graphicx}
\usepackage{amsmath}
\usepackage{amssymb}
\usepackage{xcolor}

\newcommand{\stirlingii}{\genfrac{[}{]}{0pt}{}}
\newcommand{\bea}{\begin{eqnarray}}
\newcommand{\eea}{\end{eqnarray}}

\begin{document}

\title{Non-intersecting Brownian bridges in the flat-to-flat geometry}

\date{\today}

\begin{abstract}
We study $N$ vicious Brownian bridges propagating from an initial configuration $\{a_1 < a_2 < \ldots< a_N \}$ at time $t=0$ to a final configuration 
$\{b_1 < b_2 < \ldots< b_N \}$ at time $t=t_f$, while staying non-intersecting for all $0\leq t \leq t_f$. We first show that this problem can be mapped 
to a non-intersecting Dyson's Brownian bridges with Dyson index $\beta=2$. For the latter we derive an exact effective Langevin equation that allows to
generate very efficiently the vicious bridge configurations.  
In particular, for the flat-to-flat configuration in the large $N$ limit, where $a_i = b_i = (i-1)/N$, for $i = 1, \cdots, N$, we use this effective 
Langevin equation to derive an exact Burgers' equation (in the inviscid limit)   
for the Green's function and solve this Burgers' equation for arbitrary time $0 \leq t\leq t_f$. 
At certain specific values of intermediate times $t$, such
as $t=t_f/2$, $t=t_f/3$ and $t=t_f/4$ we obtain the average density of the flat-to-flat bridge explicitly. 
We also derive explicitly how the two edges of the average density evolve from time $t=0$ to time $t=t_f$. 
Finally, we discuss connections to some well known problems, such as the Chern-Simons model, the related Stieltjes-Wigert
orthogonal polynomials and the Borodin-Muttalib ensemble of determinantal point processes.   
\end{abstract}

\author{Jacek Grela} \email{ jacekgrela@gmail.com} 
\affiliation{Institute of Theoretical Physics, Jagiellonian University, 30-348 Cracow, Poland}
\author{Satya N. Majumdar}
\email{majumdar@lptms.u-psud.fr}
\affiliation{LPTMS, CNRS, Univ. Paris-Sud, Universit\'{e} Paris-Saclay, 91405 Orsay, France}
\author{Gr\'{e}gory Schehr}
\email{gregory.schehr@u-psud.fr} 
\affiliation{Sorbonne Universit\'e, Laboratoire de Physique Th\'eorique et Hautes Energies, CNRS UMR 7589, 4 Place Jussieu, 75252 Paris Cedex 05, France}

\maketitle
\section{Introduction}

The simplest way to generate a Brownian trajectory in one dimension is to evolve the position of a particle with time, starting say from $x(0)=0$, via the stochastic Langevin equation
\begin{eqnarray}\label{Langevin_BM}
\frac{dx}{dt} = \sqrt{2 D}\, \eta(t) \;,
\end{eqnarray}
where $D$ is the diffusion constant and $\eta(t)$ is a Gaussian white noise with zero mean and two-time correlator $\langle \eta(t) \eta(t')  \rangle = \delta(t-t')$. In many practical 
situations one needs to generate a constrained Brownian motion, e.g., a Brownian bridge that starts at $x(0)=a$ and ends at $x(t_f) =b$ after a fixed time $t_f$. Naively, one
would generate all possible free Brownian configurations and retain, amongst them, only those that satisfy the bridge condition. Numerically, this is of course
extremely inefficient. In the probability literature, such conditioned Brownian motions have been studied extensively pioneered by Doob \cite{doob1957conditional,fitzsimmons1993markovian}, where one works out the transition
probability of the effective Markov process that satisfies this constraint -- known as the Doob transform. However, for numerical purposes, it is desirable to 
write explicitly an effective Langevin equation that generates constrained Brownian motions (such as the bridge)
with the correct statistical weight \cite{orland2011generating,chetrite2013nonequilibrium}. In a recent work \cite{MO2015:CONSTRAINTLANGEVIN}, it has been shown how to
construct explicitly such an effective Langevin equation for a class of constrained stochastic Markov processes in which the constraint manifests itself explicitly  
as an additional force in the Langevin equation. In the case of the Brownian bridge, this effective equation reads \cite{MO2015:CONSTRAINTLANGEVIN}
\begin{eqnarray} \label{Langevin_BB}
\frac{dx(t)}{dt} = \frac{b-x(t)}{t_f-t} + \sqrt{2 D}\, \eta(t) \;,
\end{eqnarray}
where $\eta(t)$ is the same Gaussian white noise. Note that the force-term in (\ref{Langevin_BB}) is time dependent and its presence ensures that the particle
reaches $b$ exactly at time $t_f$. The procedure used in \cite{MO2015:CONSTRAINTLANGEVIN} is rather general and has been used to derive the effective Langevin
equation for several other constrained stochastic processes, such as an Ornstein-Uhlenbeck bridge, Brownian excursion, meander, etc \cite{MO2015:CONSTRAINTLANGEVIN}.
This was done for the case of a single particle under constraints. It is then natural to wonder if one can develop a similar approach for many-body interacting Brownian particles.

A particularly simple model of interacting walkers is the so-called ``vicious walkers'' model introduced by de Gennes \cite{gennes1968soluble}, followed by Fisher \cite{fisher1984walks} and others \cite{huse1984commensurate,johansson2002non,prahofer2002scale,katori2004symmetry,tracy2007nonintersecting,schehr2008exact,borodin2009maximum,nadal2009nonintersecting,rambeau2010extremal,forrester2011non,schehr2013reunion,nguyen2017extreme,le2017periodic,gautie2019non}. There have been a lot of applications of this model both in statistical physics, for example in the context of fibrous polymers \cite{gennes1968soluble} or the melting and wetting transition in solids \cite{fisher1984walks,huse1984commensurate}, all the way to combinatorics \cite{krattenthaler2000vicious} and computer science \cite{bonichon2003watermelon}. Here one considers $N$ Brownian motions starting at $a_1< a_2< \cdots< a_N$ and their positions $\{x_1(t), x_{2}(t), \cdots, x_N(t) \}$ at time $t$ evolve via the free Langevin equation (\ref{Langevin_BM}) except that they are constrained never to cross each other. In this model, for fixed $N$, there is thus just one parameter, namely the diffusion constant $D$, 
which is assumed to be the same for all the particles. Because of the non-intersection constraints, their positions at any time $t$ remain ordered, i.e., $x_1(t)<x_2(t)<\cdots < x_N(t)$, provided they are ordered initially.

A natural practical question is: how to generate numerically a bridge configuration for such vicious Brownian motions that start at $a_1< a_2< \cdots< a_N$ at $t=0$ and end at $b_1< b_2< \cdots< b_N$ at time $t=t_f$. Such vicious Brownian
bridges (VBB) are known in the literature as ``watermelons'' (because they look like one) and they appear naturally 
in both physics and computer science. Finding an efficient algorithm to generate a watermelon configuration is a
challenging problem \cite{bonichon2003watermelon}. A naive answer to this question would be to generate configurations of $N$ free Brownian bridges, up to time $t$, and then retain only those where the positions do not cross each other during the interval $[0,t_f]$. But this obviously is a rather wasteful algorithm.

In this paper, generalising the formalism of Ref. \cite{MO2015:CONSTRAINTLANGEVIN} to $N$ non-intersecting
Brownian motions, we provide an exact algorithm to generate a VBB. We will proceed in two steps: (i) First we make an exact mapping between 
the VBB with diffusion constant $D = 1/(2N)$ and the so-called Dyson Brownian Bridge (DBB) with Dyson index $\beta = 2$ (for a precise definition of the DBB see Section \ref{sec1}). We show that under this mapping the positions of the $N$ walkers in the VBB model, at any intermediate time $0\leq t \leq t_f$, coincide in law with the positions of the $N$ particles of DBB (with $\beta = 2$) at the same time $t$. By ``coincide in law'' one means that the joint distribution of the positions at any intermediate time  are identical in both models. Therefore, if we can generate numerically  the configurations of the DBB (with $\beta = 2$), this will automatically generate a configuration of the VBB model. (ii) Next we show that one can write an exact effective Langevin equation to generate the configurations of the DBB  
with $\beta = 2$. In particular, when the final positions are uniformly distributed over the interval $[0,1]$, 
obtained by choosing $b_i = (i-1)/N$, this effective Langevin equation becomes totally explicit. 
Thus (i) and (ii) together provide us with an exact algorithm to generate the configurations of the VBB.

In addition to providing a numerically efficient algorithm to generate a watermelon configuration, we show that our explicit Langevin equation, for the case
when both the initial and the final configurations are uniformly distributed (``flat to flat''), i.e. $a_i =b_i= (i-1)/N$, with $i=1,2, \cdots, N$, 
can be successfully exploited analytically to compute certain observables in the large $N$ limit, such as the average density $\rho(\lambda;t)$ of the walkers at any intermediate time
in the interval $[0,t_f]$. In particular, we provide explicit formulae for $\rho(\lambda;t)$ at $t=t_f/4$, $t=t_f/3$, $t=t_f/2$ -- note that the average density
is symmetric around $t=t_f/2$ for the flat-to-flat case. Hence the density at $t$ and $t_f-t$ are identical. We note that the previous result was known, by completely
different methods, only for $t=t_f/2$. Indeed this particular case $t=t_f/2$ appeared in two apparently unrelated contexts, namely the Chern-Simons theory \cite{marino2005chern,Mar2006:MATRIXMODELSSTRINGS}
and the theory of Stieltjes-Wigert orthogonal polynomials \cite{DT2007:SWANDCS,szabo2010chern,TK2012:NONCOLLSW,borot2016asymptotic}. In fact, this problem also has some connections to the
computation of the Harish-Chandra-Itzykson-Zuber (HCIZ) integral with flat initial and final configurations of the eigenvalues of two matrices $A$ and $B$ \cite{guionnet2004first,BBMP2014:HCIZINTEGR,menon2017complex}. 
For other values of $t$, determining $\rho(\lambda;t)$ explicitly is hard. However, one can show that $\rho(\lambda;t)$ 
has a finite support $[\lambda_-(t), \lambda_+(t)]$ and we are able to compute explicitly how the support edges $\lambda_{\pm}(t)$ evolve with time [see Eq.~\eqref{endpointslambda}].

The rest of the paper is organised as follows. In Section \ref{sec1} we provide (i) the mapping between the VBB and the DBB with $\beta = 2$ and (ii) derive the exact effective Langevin equations for the $N$-particle DBB for $\beta = 2$ where the final positions are equi-spaced (flat density). Next, in Section \ref{sec2}, we provide a derivation of the evolution equation of the Green's function and show that, in the large $N$ limit, it reduces to the inviscid Burgers' equation, which can then be solved using the method of characteristics. In Section \ref{sec3}, we focus on the case where both the initial and the final density are flat and we compute explicitly the large $N$ average density for three specific values of the time instants, namely $t=t_f/4, t_f/3$ and $t=t_f/2$, and these are detailed in subsections IV.A, IV.B and IV.C respectively. In Subsection IV.D, we derive explicitly the time evolution $\lambda_{\pm}(t)$ of the edges of the support of the average density $\rho(\lambda;t)$. We also provide in Section IV.E a recursive derivation of the moments of the average density. In Section \ref{sec4} we 
relate the flat-to-flat VBB to other models such as the Chern-Simons model and the theory of the biorthogonal Stieltjes-Wigert polynomials appearing in  
Muttalib-Borodin ensembles. Finally, we conclude in Section VI with a summary and outlook.

\section{The mapping between the VBB and DBB and an effective Langevin equation for the DBB}
\label{sec1}

\subsection{The exact equivalence between the VBB and DBB with $\beta = 2$}

\begin{figure}[t]
\includegraphics[width = 0.4\linewidth]{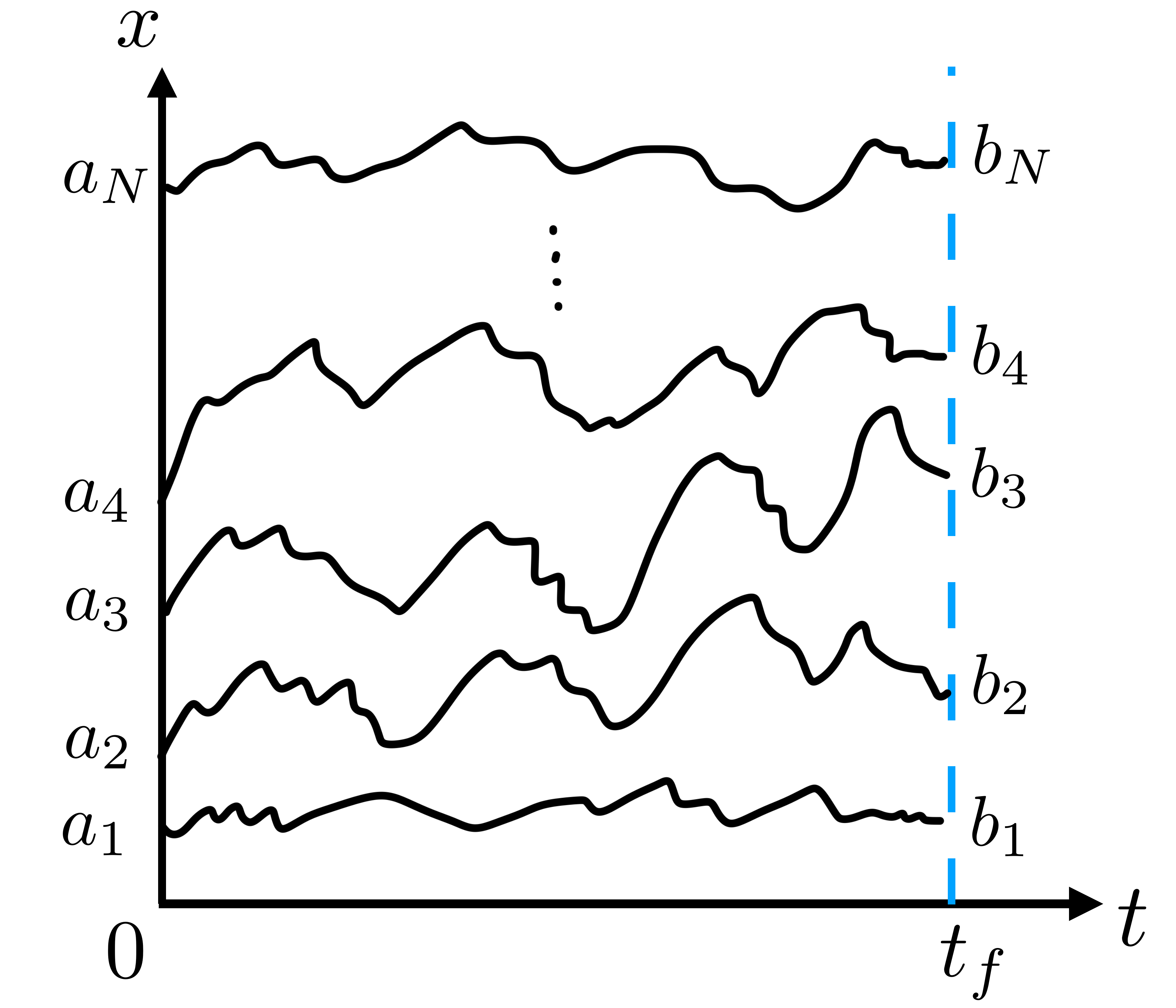}
\caption{Sketch of the trajectories of a VBB starting form ${\vec a}$ at time $t=0$ and ending at $\vec{b}$ at time $t_f$.}\label{Fig_VBB}
\end{figure}

We start with the vicious walkers bridge (VBB) starting at $t=0$ from $\vec{a} = \{a_1, a_2, \cdots, a_N\}$ with $a_1<a_2< \cdots < a_N$, and
ending at $t=t_f$ at $\vec{b} = \{b_1, b_2, \cdots, b_N\}$ with $b_1<b_2< \cdots < b_N$ (see Fig. \ref{Fig_VBB}). Let $\vec{\lambda}(t) = \{ \lambda_1(t), \lambda_2(t), \cdots, \lambda_N(t)$ denote the positions of the $N$ walkers in the VBB at an intermediate time $t$. Each of the $\lambda_i$'s evolves locally in time via the Langevin equation 
\bea \label{Langevin_lamb}
\frac{d\lambda_i}{dt} = \sqrt{2\,D} \, \eta_i(t) \;,
\eea 
where $\eta_i$'s are again Gaussian white noises with zero mean and correlator $\langle \eta_i(t) \eta_j(t')\rangle = \delta_{i,j} \delta(t-t')$. Here $D$ is the diffusion constant that we assume to be the same for each walker. To compute the joint distribution 
$P_{{\rm VBB},D}(\vec{\lambda}, t|\vec{b},\vec {a},t_f)$ of these positions at time $t$, given the initial and final positions, we proceed as follows. Dividing the time interval $[0,t_f]$ into $[0,t]$ and $[t,t_f]$ with $0 \leq t \leq t_f$ and using the Markov property of the process, we get  
\begin{eqnarray}\label{P_vicious_bridge}
P_{{\rm VBB},D}(\vec{\lambda}, t|\vec{b},\vec {a},t_f) = \frac{P_{{\rm VBM},D}(\vec{\lambda}, t | \vec{a},0)\, P_{{\rm VBM},D}(\vec{b}, t_f | \vec{\lambda},t)}{P_{{\rm VBM},D}(\vec{b}, t_f | \vec{a},0)}  \;,
\end{eqnarray}
where $P_{{\rm VBM},D}(\vec{\lambda}, t | \vec{a},0)$ denotes the propagator from time $0$ to time $t$ of the free vicious Brownian motion (VBM) -- without the bridge constraint. This means the propagator for $N$ non-intersecting Brownian motions starting from $\vec{a}$  
at $t=0$ to $\vec{\lambda}$ at time $t$. We then need to compute this propagator for VBM, as an intermediate step, to compute the joint distribution of the VBB via Eq. (\ref{P_vicious_bridge}).  We now show that this propagator $P_{{\rm VBM},D}(\vec{\lambda}, t | \vec{a},0)$ can be related to the propagator of the so-called Dyson Brownian motion (DBM) with $\beta = 2$.

Let us first recall the definition of the DBM with Dyson index $\beta$. In DBM, one again considers $N$ particles on a line, with positions    
$\lambda_i(t)$ with $i=1, 2, \cdots, N$, that evolve via the stochastic equation 
\begin{align}
\label{model}
\frac{d\lambda_i(t)}{dt} = \frac{1}{N} \sum_{j(\neq i)=1}^N\frac{1}{\lambda_i(t)-\lambda_j(t)}  + \sqrt{\frac{2}{\beta\,N}}\;\xi_i(t)\;, 
\end{align}
where $\xi_i(t)$ are zero mean Gaussian white noises with correlations  $\left < \xi_i(t) \xi_j(t')\right > =  \delta_{ij} \delta(t-t')$. The DBM, for fixed $N$, has only one parameter $\beta$, known as the Dyson's index. This model comes from random matrix theory (RMT) where one considers 
$N \times N$ random matrices (real symmetric, complex Hermitian or complex quaternionic) where the entries undergo independent Brownian motions as in Eq. (\ref{Langevin_BM}). 
At each time $t$, if one diagonalises the matrix, one gets $N$ real eigenvalues $\{\lambda_1(t)< \lambda_2(t) < \cdots < \lambda_N(t) \}$. Dyson, using second order perturbation theory, demonstrated \cite{Dys1962:BROWNIAN} that the eigenvalues evolve via the stochastic equations (\ref{model}). The parameter $\beta = 1, 2$ and $4$ correspond to the real symmetric, complex Hermitian or complex quaternionic matrices. The force term on the right hand side of Eq. (\ref{model}) comes from the effective pairwise repulsion between the eigenvalues. 
Even though originally only three quantized values of $\beta = 1, 2, 4$ were studied due to their relation with the symmetry classes of Gaussian matrices, it was later shown that there are actually
matrix models which give rise to this Langevin equation (\ref{model}) for arbitrary $\beta  > 0$ \cite{dumitriu2002matrix}.  The trajectories of the DBM, for arbitrary $\beta>0$,
can be easily generated numerically by evolving the positions of the walkers according to Eq. (\ref{model}) starting from ordered initial positions $a_1< a_2< \cdots< a_N$. What is the connection between the two models: (i) VBM with parameter $D$ and (ii) DBM with parameter $\beta$?

To see this connection, we first note that the propagator in the VBM with diffusion constant $D$ satisfies the Fokker-Planck equation in the region $\lambda_1 \leq \lambda_2 \leq \cdots \leq \lambda_N$
\begin{eqnarray}\label{FP_VBM}
\frac{\partial P_{\rm VBM}(\vec{\lambda},t)}{\partial t}= D \sum_{i=1}^N \frac{\partial^2 P_{\rm VBM}(\vec{\lambda},t)}{\partial \lambda_i^2} \;,
\end{eqnarray}
starting from the initial condition $P_{\rm VBM}(\vec{\lambda},t=0) = \delta(\vec{\lambda} - \vec{a})$. Note that, for simplicity of notations, we have omitted the explicit dependence of $P_{\rm VBM}(\vec{\lambda},t)$ on $\vec{a}$ in Eq. (\ref{FP_VBM}). In addition, $P_{\rm VBM}(\vec{\lambda},t)$ satisfies the boundary conditions 
\begin{eqnarray}\label{BC_VBM}
P_{\rm VBM}(\lambda_i = \lambda_j,t) = 0 \quad \forall \; i \neq j \;.
\end{eqnarray}
These boundary conditions ensure the non-intersection constraint. 

Now consider the DBM in Eq. (\ref{model}). The associated Fokker-Planck equation in the region $\lambda_1 < \lambda_2 < \cdots < \lambda_N$ evolves as
\begin{eqnarray} \label{FP_DBM}
\frac{\partial P_{\rm DBM}(\vec{\lambda},t)}{\partial t} = \frac{1}{\beta\,N} \sum_{i=1}^N \frac{\partial^2 P_{\rm DBM}(\vec{\lambda},t)}{\partial \lambda_i^2} -\sum_{i=1}^N \frac{\partial}{\partial \lambda_i} \left( E_i(\vec{\lambda}) P_{\rm DBM}(\vec{\lambda},t) \right) \quad; \quad {\rm where} \quad E_i(\vec{\lambda}) = \frac{1}{N} \sum_{j (\neq i)=1}^N \frac{1}{\lambda_i - \lambda_j} \;,
\end{eqnarray}
with the initial condition $P_{\rm DBM}(\vec{\lambda},t=0) = \delta(\vec{\lambda} - \vec{a})$. The explicit repulsion term in (\ref{FP_DBM}) automatically ensures that the propagator $P_{\rm DBM}(\vec{\lambda},t)$ vanishes whenever $\lambda_i = \lambda_j$ with $i \neq j$ as in Eq. (\ref{BC_VBM}). To relate this propagator to that of the VBM, we first make the transformation
\begin{eqnarray}\label{transfo_FP}
P_{\rm DBM}(\vec{\lambda},t) = \left( \frac{\Delta(\vec{\lambda})}{\Delta(\vec{a})}\right)^{\beta/2}\; W(\vec{\lambda},t) \;,
\end{eqnarray}  
where $\Delta(\vec{\lambda}) = \prod_{i<j}(\lambda_j - \lambda_i)$ and similarly $\Delta(\vec{a}) = \prod_{i<j}(a_j - a_i)$ are Vandermonde determinants. 
Upon substituting this form (\ref{transfo_FP}) in Eq. (\ref{FP_DBM}) and after long but straightforward algebra one finds \cite{rambeau2011distribution} that $W(\vec{\lambda},t)$ satisfies
\begin{eqnarray}\label{FP_W}
\frac{\partial W(\vec{\lambda},t)}{\partial t} =  \frac{1}{\beta\,N} \sum_{i=1}^N \frac{\partial^2 W(\vec{\lambda},t)}{\partial \lambda_i^2} - \,\frac{\beta-2}{2N} \sum_{i=1}^N \sum_{1 \leq j\neq i \leq N} \frac{1}{(\lambda_i-\lambda_j)^2} \, W(\vec{\lambda},t) \;.
\end{eqnarray}
For general $\beta$, $W(\vec{\lambda},t)$ can be interpreted as the imaginary time propagator for the quantum Calogero-Sutherland model. Note that for $\beta = 2$ the second term on the right hand side (rhs) of Eq. (\ref{FP_W}) vanishes and $W({\vec \lambda},t)$ then satisfies the same equation as (\ref{FP_VBM}) with $D=1/(2N)$, with the same boundary and initial conditions. Hence, for $\beta = 2$ and $D=1/(2N)$, using Eq. (\ref{transfo_FP}), we get the identity   
 \begin{eqnarray}\label{relation_vicious}
 P_{{\rm DBM}, \beta=2}(\vec{\lambda},t|\vec{a},0) = \frac{\Delta(\vec{\lambda})}{\Delta(\vec{a})} \, P_{{\rm VBM}, D=\frac{1}{2N}}(\vec{\lambda},t|\vec{a},0) \;.
 \end{eqnarray}

We now consider a Dyson Brownian bridge (DBB) where the positions of $N$ particles in the DBM (\ref{model}) starting at $\vec{a}$ at time $t=0$ are further constrained to reach $\vec{b}$ at time $t=t_f$. As was done for the case of the VBB in (\ref{P_vicious_bridge}), one can again use the Markov property of the DBM to write the joint distribution $P_{{\rm DBB},D}(\vec{\lambda}, t|\vec{b},\vec {a},t_f)$ of these positions at time $t$, given the initial and final positions as
\begin{eqnarray}\label{P_DBB}
P_{{\rm DBB}, \beta}(\vec{\lambda}, t|\vec{b},\vec {a},t_f) = \frac{P_{{\rm DBM}, \beta}(\vec{\lambda}, t | \vec{a},0)\, P_{{\rm DBM}, \beta}(\vec{b}, t_f | \vec{\lambda},t)}{P_{{\rm DBM}, \beta}(\vec{b}, t_f | \vec{a},0)}  \;.
\end{eqnarray}
Setting $\beta=2$ and $D=1/(2N)$, and plugging the relation (\ref{relation_vicious}) in Eqs. (\ref{P_vicious_bridge}) and (\ref{P_DBB}), we see that 
the Vandermonde terms $\Delta$'s cancel out giving us the exact relation
\begin{eqnarray}\label{equivalence}
P_{{\rm VBB}, D =\frac{1}{2N}}(\vec{\lambda}, t|\vec{b},\vec {a},t_f)  = P_{{\rm DBB}, \beta=2}(\vec{\lambda}, t|\vec{b},\vec {a},t_f) \;.
\end{eqnarray}
Note that this relation is valid for any time $0\leq t \leq t_f$. This implies that the positions of the $N$ walkers at any intermediate time $t$ have
the same statistics in the two models. Statistically speaking, this means that 
\begin{eqnarray}\label{id_positions}
\{ \lambda_1(t), \lambda_2(t), \cdots, \lambda_N(t)\}_{{\rm VBB}, D = \frac{1}{2N}} \equiv  \{ \lambda_1(t), \lambda_2(t), \cdots, \lambda_N(t)\}_{\rm DBB, \beta=2}
\end{eqnarray}
where the symbol ``$\equiv$'' indicates equivalence in law. This relation tells us that if we know how to generate numerically a configuration of the DBB with $\beta=2$, then that configuration will also be a configuration of the VBB with $D = 1/(2N)$. Note that this choice of $D=1/(2N)$ is not restrictive at all since one can always rescale time in the original Langevin equation (\ref{Langevin_lamb}) to set
the diffusion constant to any prescribed value. Thus our mapping constitutes the first step towards building an algorithm to generate a VBB configuration. The second step is to generate explicitly a DBB configuration 
with $\beta = 2$. For this, we will generalize the method developed in
Ref. \cite{MO2015:CONSTRAINTLANGEVIN} to write an effective Langevin equation to generate a DBB with $\beta=2$. This is done in the next subsection.

\subsection{Effective Langevin equation for the DBB with $\beta=2$}

In this subsection, we generalize the method of Ref. \cite{MO2015:CONSTRAINTLANGEVIN} to the case of a DBB of $N$ particles and arbitrary positive index $\beta >0$.
The positions of the particles $\vec{\lambda}(t)$ evolve via the Langevin equations (\ref{model}), starting from the initial positions $\vec{a}$. The joint distribution of the
particle positions at any time $t$ is given in Eq. (\ref{P_DBB}). For convenience, we denote the second term in the numerator as
\begin{eqnarray}\label{def_Q}
P_{{\rm DBM}, \beta}(\vec{b}, t_f | \vec{\lambda},t) = Q(\vec{\lambda},t | \vec{b}, t_f) \;.
\end{eqnarray}
Note that the left hand side (lhs) is the ``forward'' propagator from $\vec{\lambda}$ at time $t$ to ${\vec{b}}$ at time $t_f$, while the rhs is the same
quantity but interpreted as the ``backward'' (i.e., time reversed) propagator from $\vec{b}$ at time $t_f$ to $\vec{\lambda}$ at time $t$. For convenience, we rewrite the
numerator in (\ref{P_DBB}) in a shorthand notation as
\begin{eqnarray}\label{P_DBB_short}
\tilde P(\vec{\lambda},t) = P_{{\rm DBM}, \beta}(\vec{\lambda},t|\vec{a},0)\, Q(\vec{\lambda},t | \vec{b}, t_f) \;.
\end{eqnarray}
The goal is to derive a Fokker-Planck (FP) equation for $\tilde P(\vec{\lambda},t)$, knowing the FP equations for $P_{{\rm DBM}, \beta}$ and $Q(\vec{\lambda},t | \vec{b}, t_f)$. 
Using further shortcut notations $P \equiv P_{{\rm DBM}, \beta}$ and $Q \equiv Q(\vec{\lambda},t | \vec{b}, t_f)$, it is easy to see, given the Langevin equation (\ref{model}), that $P$ and $Q$ satisfy respectively the forward and backward FP equations 
\begin{align}
\partial_t P & = \tilde D \sum_{i=1}^N \partial_{\lambda_i}^2 P - \sum_{i=1}^N \partial_{\lambda_i} \left [ E_i(\vec \lambda)\, P \right ], \label{forwardeq} \\
- \partial_t Q & = \tilde D \sum_{i=1}^N \partial_{\lambda_i}^2 Q + \sum_{i=1}^N E_i(\vec{\lambda}) \, \partial_{\lambda_i} Q , \label{backsfp}
\end{align}
where 
\begin{equation}\label{def_Dtilde}
\tilde D = \frac{1}{\beta N} \quad\quad, \quad \quad  E_i(\vec{\lambda}) = \frac{1}{N} \sum\limits_{j (\neq i) = 1}^N \frac{1}{\lambda_i - \lambda_j} \;.
\end{equation}
Note that in the equation for $Q$ (\ref{backsfp}) on the lhs the time derivative has a negative sign. This is because the time is going backward. Since $\tilde P = P\, Q$, one gets

 \begin{align} \label{Ptilde_1}
\partial_t \tilde{P} = Q \,\partial_t P + P \, \partial_t Q = \tilde{D} \sum_i \left [ Q \, \partial^2_{\lambda_i} P - P \partial^2_{\lambda_i} Q \right ] -  \sum_i \left [ Q \, \partial_{\lambda_i} \left ( E_i(\vec \lambda)\, P \right ) + P  \, E_i(\vec{\lambda})\, \partial_{\lambda_i} Q  \right ].
\end{align}
We next use the following two identities
\begin{eqnarray}
&&Q \partial^2_{\lambda_i} P - P \partial^2_{\lambda_i} Q = \partial^2_{\lambda_i}\tilde {P} - 2 \partial_{\lambda_i} \left ( \tilde{P} \, \partial_{\lambda_i} \log Q \right ) \;, \label{id1}\\
&&Q \partial_{\lambda_i} \left ( E_i(\vec{\lambda}) \, P \right ) + P \, E_i(\vec{\lambda})\, \partial_{\lambda_i} \, Q  = \partial_{\lambda_i} \left ( E_i(\vec{\lambda}) \, \tilde{P} \right) \;,\label{id2}
\end{eqnarray}
to simplify the rhs of Eq. (\ref{Ptilde_1}) and obtain a compact FP equation for $\tilde{P}$ 
\begin{align}
\label{ptilde}
\partial_t \tilde{P} = \tilde D \sum_i \partial^2_{\lambda_i} \tilde{P} - \sum_i \partial_{\lambda_i} \left [ \left ( E_i(\vec{\lambda}) + 2 \,\tilde D\, \partial_{\lambda_i} \log Q \right ) \tilde{P} \right ] \;.
\end{align}
This FP equation corresponds exactly to an effective Langevin equation
\begin{eqnarray}\label{Lang_eff}
\frac{d\, \lambda_i(t)}{dt} = E_i(\vec{\lambda}) + 2 \, \tilde D \, \frac{\partial \ln{Q}}{\partial \lambda_i} + \sqrt{2 \, \tilde D}\, \xi_i(t) \;,
\end{eqnarray}
where $\xi_i(t)$ is a Gaussian white noise with zero mean and correlator $\langle \xi_i(t) \xi_j(t') \rangle = \delta_{ij} \delta(t-t')$. Using the expressions for $\tilde D$ and $E_i$ in Eq. (\ref{def_Dtilde}), we finally get the effective Langevin equation of the DBB with arbitrary $\beta > 0$ as
\begin{eqnarray}\label{Lang_eff2}
\frac{d\, \lambda_i(t)}{dt} = \frac{1}{N}\, \sum_{j \neq i} \frac{1}{\lambda_i-\lambda_j} + \frac{2}{\beta N} \, \frac{\partial \ln{Q}}{\partial \lambda_i} + \sqrt{\frac{2}{\beta N}}\; \xi_i(t) \;.
\end{eqnarray}
Note that in this equation the second term represents the effective force due to the bridge constraint. Of course, one needs to know the backward propagator $Q$ for the free DBM in order to compute this effective force term explicitly. For a general $\beta$, this is very hard. However, for $\beta=2$, using the connection with the VBM, one can make progress. Indeed, in this case, using the relation (\ref{relation_vicious}), we get 
\begin{eqnarray}\label{KMG1}
Q(\vec{\lambda},t| \vec{b}, t_f) = P_{{\rm DBM}, \beta = 2}(\vec{b},t_f|\vec{\lambda},t) = \frac{\Delta(\vec{b})}{\Delta(\vec{\lambda})} P_{{\rm VBM},D = \frac{1}{2N}}(\vec{b},t_f|\vec{\lambda}, t) \;. 
\end{eqnarray}
It turns out that the propagator for the VBM can be computed explicitly using the Karlin-McGregor formula \cite{KMcG1959:KARLINMCGREGORFORMULA1}
\begin{eqnarray}\label{KMG2}
P_{{\rm VBM},D}(\vec{b},t_f|\vec{\lambda}, t) = \det_{1 \leq i,j \leq N} \left[ \frac{e^{-\frac{(b_i-\lambda_j)^2}{4\,D\,(t_f-t)}}}{\sqrt{4 \pi D\,(t_f-t)}}\right] \;.
\end{eqnarray} 
In general, it is not easy to evaluate this $N \times N$ determinant explicitly for arbitrary $b_i$'s. However, for the special case where the final positions are
equispaced (corresponding to a flat density over the interval $\lambda \in [0,1]$)
\begin{eqnarray} \label{equi}
b_i = \frac{i-1}{N} \;, \; i  =1, 2, \cdots, N \;,
\end{eqnarray}  
this determinant can be explicitly evaluated as follows. Expanding the determinant in (\ref{KMG2}) we get
\bea \label{expand_KMG}
 \det_{1 \leq i,j \leq N} \left ( \frac{e^{-\frac{(b_i - \lambda_j)^2}{4D(t_f-t)}}}{\sqrt{4\pi D(t_f -t)}} \right ) =  \frac{e^{-\frac{1}{4D(t_f-t)}\sum_j \lambda_j^2} e^{-\frac{1}{4D(t_f-t)}\sum_j b_j^2}}{\left [ 4\pi D(t_f -t) \right ]^{N/2}}  \det_{1 \leq i,j \leq N}\left ( e^{\frac{b_i \lambda_j}{2D(t_f-t)}} \right ) \;.
\eea
For the flat final condition (\ref{equi}) the determinant in \eqref{expand_KMG} can be further simplified
\begin{eqnarray}\label{expand_KMG2}
\det_{1 \leq i,j \leq N} \left ( e^{\frac{b_i \lambda_j}{2D(t_f-t)}} \right ) = \det_{1 \leq i,j \leq N} \left ( e^{\frac{ (i-1)\,\lambda_j}{2D\,N(t_f-t)}} \right ) =
\prod_{i<j} \left ( e^{\frac{\lambda_j}{2DN(t_f-t)} } - e^{\frac{\lambda_i}{2DN(t_f-t)} } \right ),
\end{eqnarray}
where we have used the fact that $\det_{1 \leq i,j \leq N} X_j^{i-1} = \prod_{i<j}(X_j - X_i)$. This gives
\begin{eqnarray}\label{KMG3}
P_{{\rm VBM},D}(\vec{b},t_f|\vec{\lambda}, t) \propto e^{-\frac{1}{4\,D\,(t_f-t)} \sum_{i=1}^N \lambda_i^2} \prod_{i<j} \left(e^{\frac{\lambda_j}{2DN(t_f-t)}} - e^{\frac{\lambda_i}{2DN(t_f-t)}}\right) \;,
\end{eqnarray}
where the proportionality constant (omitted here) is independent of $\lambda$ and does not play any role (as we will see shortly). Inserting this result in Eq.~(\ref{KMG1}), and taking logarithm and using $D = 1/(\beta N)= 1/(2N)$ for $\beta = 2$, gives 
\begin{eqnarray}\label{Q_explicit}
\ln Q = - \frac{N}{2(t_f-t)}\sum_{i=1}^N \lambda_i^2 - \sum_{i<j} \ln(\lambda_j-\lambda_i) + \sum_{i<j} \ln \left( e^{\frac{\lambda_j}{t_f-t}} - e^{\frac{\lambda_i}{t_f-t}}\right) + A \;,
\end{eqnarray}
where $A$ is a constant which is independent of $\vec{\lambda}$. Taking derivative of (\ref{Q_explicit}) with respect to $\lambda_i$ and inserting in (\ref{Lang_eff2}) we get an explicit Langevin equation
\begin{align}
\label{Langevinlambda}
\frac{d\lambda_i}{dt} = -\frac{\lambda_i }{t_f-t} + \frac{1}{N(t_f-t)} \sum_{j(\neq i)} \frac{e^{\frac{\lambda_i}{t_f-t} }}{e^{\frac{\lambda_i}{t_f-t} } - e^{\frac{\lambda_j}{t_f-t} }} + \frac{1}{\sqrt{N}}\, \xi_i(t) \;.
\end{align}
This equation starts from any arbitrary initial condition $\vec{a}$. It does generate the trajectories of the DBB with $\beta = 2$ (and hence of the VBB with $D = 1/(2N)$) where the final positions correspond to a flat density in Eq. (\ref{equi}). This completes our derivation of the Langevin equation for the VBB. Numerically it is easy to generate the trajectories of VBB using this equation. Examples of such trajectories are shown in Figs. \ref{fig1}, \ref{fig2} and \ref{fig3}.

By examining the rhs of Eq. (\ref{Langevinlambda}), it is natural to make the following change of variables
\begin{align}\label{change}
\begin{cases}
&x_i = e^{\frac{\lambda_i}{t_f-t}} \;, \\
& \\
&\theta = \frac{t}{t_f-t} \;,
\end{cases}
\end{align}
or equivalently  
\begin{align}
\begin{cases}
&\lambda_i = \frac{t_f}{1+\theta} \log x_i \;, \\ 
& \\
&t = t_f \frac{\theta}{1+\theta} \;,
\end{cases}
\label{transformation}
\end{align}
where the new space-- and time--like variables become positive $x_i \in (0,\infty) $ and $\theta \in (0,\infty)$. Since $\theta = t/(t_f-t)$, the initial time $t=0$ is mapped to $\theta=0$ while the final time $t=t_f$ corresponds to $\theta \to \infty$. The Langevin equations (\ref{Langevinlambda}) in terms of these new coordinates read (after some straightforward algebra)
\begin{align}
\label{Langevinx}
\frac{dx_i}{d\theta} = \frac{1}{N t_f} \sum_{j(\neq i)} \frac{x_i^2}{x_i - x_j} + \frac{1+\theta}{t_f} x_i \, \tilde \xi_i(\theta) \;,
\end{align}
where $\tilde \xi_i(\theta)$ is a Gaussian white noise with zero mean and correlator $\langle\tilde  \xi_i(\theta) \tilde \xi_j(\theta') \rangle= \delta_{ij} \delta(\theta - \theta')$. Note that the drift term in this transformed Langevin equation (\ref{Langevinx}) does not contain explicit $\theta$-dependence, unlike the original Eq. (\ref{Langevinlambda}) where the drift term depends on $t$ explicitly. In addition, the noise term in (\ref{Langevinx}) is multiplicative and we will interpret it in the Ito sense.

\section{Average particle density via Burgers' equation}
\label{sec2}

In the previous section, we have shown how to generate numerically the configuration of a VBB, by generating a configuration of DBB with
$\beta=2$ via the explicit Langevin equation (\ref{Langevinlambda}) for the case of a flat final density at $t=t_f$. In this section, we show that this effective Langevin equation is useful not just to generate a configuration numerically, but also to compute some physical observables, such as the density of the particles at an intermediate time $0\leq t \leq t_f$. 

\subsection{Derivation of the Burgers' equation}

In this section we first discuss densities in both spaces $(\lambda,t)$ [see Eq. (\ref{Langevinlambda})] and $(x,\theta)$ [see Eq. (\ref{Langevinx})] and a functional relation between them. We then introduce the Green's function as an intermediate tool to compute the average densities. To derive an evolution equation for the Green's function, it turns out to be convenient first to write an evolution equation for an object, 
analog of a characteristic polynomial. Using a Cole-Hopf relation between the characteristic polynomial and the Green's function, we can then derive an equation for the Green's function. It turns out that in the large $N$ limit this evolution equation for the Green's function simplifies a lot allowing us to obtain certain explicit results.

We consider the positions $\vec{\lambda}(t)$ evolving via the Langevin equation (\ref{Langevinlambda}) 
and define the density at time $t$
\begin{align} \label{def_density_lambda}
\rho^{(\lambda)}_N(\lambda;t) = \frac{1}{N} \left < \sum_{i=1}^N \delta (\lambda - \lambda_i(t)) \right > \;,
\end{align}
where $\langle \cdots \rangle$ denotes an average over the random variables $\vec{\lambda}(t)$ in \eqref{Langevinlambda} and the superscript $(\lambda)$ refers to the
density in the $\lambda$-space. From now on, we consider all quantities with a subscript $N$ as exact for any $N$ whereas the corresponding asymptotic quantities lack the subscript $\rho^{(\lambda)} = \lim\limits_{N\to \infty} \rho^{(\lambda)}_N$. Following the space-time transformation \eqref{transformation}, the density in $(x,\theta)$ variables reads
\begin{align}
\label{rhox}
\rho^{(x)}_N(x;\theta) = \frac{1}{N} \left < \sum_{i=1}^N \delta \left (x - x_i(\theta) \right ) \right > \;.
\end{align}
Since the transformation (\ref{transformation}) is bijective, the relation between the two densities is straightforward
\begin{align}
\label{rel}
\rho_N^{(\lambda)}(\lambda;t) = \frac{e^{\frac{\lambda}{t_f-t} }}{t_f - t} \rho_N^{(x)} \left ( x = e^{ \frac{\lambda}{t_f-t}};\theta = \frac{t}{t_f-t} \right ).
\end{align}
One way of obtaining $\rho^{(x)}_N$ is by considering an associated Green's function (or resolvent)
\begin{align}
\label{greenx}
G^{(x)}_N(y;\theta) = \frac{e^y}{N t_f} \left < \sum_{i=1}^N \frac{1}{e^y-x_i(\theta)} \right > \;,
\end{align}
which is defined on the whole complex $y$-plane with exception of the positions $e^y=x_i$. In our case, it is a real axis half--line $e^y>0$ since $x_i>0$. Knowing the Green's function, one can extract the average density using the Sochocki-Plemelj formula 
\begin{eqnarray}\label{SP}
\rho^{(x)}_N = \frac{t_f}{\pi} \lim\limits_{\epsilon \to 0_+} \text{Im} \left [  \frac{1}{y} G^{(x)}_N(\ln y;\theta) \right ]_{y = x-i\epsilon} \;,
\end{eqnarray}
where ${\rm Im}(z)$ denotes the imaginary part of $z$. Because we have already introduced the variables $(x,\theta)$ to ease the calculations, we do not define the Green's function corresponding to the $(\lambda,t)$ space and only use relation \eqref{rel} when needed.

Our goal next is to write an evolution equation for the Green's function $G^{(x)}_N(y;\theta)$. This however turns out to be hard for finite $N$. To make progress, we instead introduce 
the \emph{characteristic polynomial}
\begin{align}
\label{chardet}
\Omega_N(y;\theta) = \left < \prod_{i=1}^N \left (e^y - x_i(\theta) \right ) \right > \;,
\end{align}
where again we choose the usual argument to be in an exponential form $e^y$ for convenience. It turns out that, starting from the Langevin equation (\ref{Langevinx}), one can derive an exact evolution equation for $\Omega_N(y;\theta)$ for any finite $N$. The derivation is detailed in Appendix \ref{AppB}. This equation reads
\begin{align}
\label{chardeteq}
\partial_\theta \Omega_N = \frac{N-1}{2t_f} \Omega_N + \frac{1}{2Nt_f} \left ( \partial_y \Omega_N - \partial_{yy} \Omega_N \right ) \;.
\end{align}
Let us remark that the prefactor of the second derivative in the rhs is negative, i.e., corresponding to a negative diffusion constant $-1/(2N t_f)$. We next define the logarithmic derivative  
\begin{eqnarray}\label{def_gN}
g_N(y;\theta) = \frac{1}{N t_f}\partial_y \log \Omega_N(y;\theta) = \frac{1}{N t_f}\partial_y \ln \left < \prod_{i=1}^N \left (e^y - x_i(\theta) \right ) \right>\;.
\end{eqnarray}
The evolution equation (\ref{chardeteq}) in terms of $g_N(y;\theta)$ reads
\begin{align}
\label{Burgersfinite}
\partial_\theta g_N = \frac{1}{2Nt_f} \left ( \partial_y g_N - \partial_{yy} g_N \right ) - g_N \partial_y g_N \;,
\end{align}
where we used the two identities $\partial_y \Omega_N = N t_f g_N \Omega_N$ and $\partial_{yy} \Omega_N = N t_f \Omega_N \partial_y g_N + (N t_f g_N)^2 \Omega_N$.

How do we relate the function $g_N(y;\theta)$ in Eq. (\ref{def_gN}) and the Green's function $G^{(x)}_N(y;\theta)$ defined in Eq. (\ref{greenx})? For any finite $N$ these two functions are a priori different. Indeed, $G^{(x)}_N(y;\theta)$ in Eq. (\ref{greenx}) can be expressed as 
\begin{eqnarray}\label{avg_log}
G^{(x)}_N(y;\theta) = \frac{1}{N\, t_f} \partial_y \left< \ln \prod_{i=1}^N \left(e^y - x_i(\theta) \right) \right> \;.
\end{eqnarray}
Comparing this expression to that of $g_N(y;\theta)$ in (\ref{def_gN}), we see that while in Eq. (\ref{def_gN}) the average $\langle \cdots \rangle$ 
is inside the argument of the logarithm, it is outside the logarithm in Eq. (\ref{avg_log}). However, typically in the limit of large $N$, the random variable inside the logarithm is highly peaked and the two averages coincide in that limit -- this is known as the ``self-averaging property''. If the self-averaging property holds (which we will assume here), we then have 
\begin{align} \label{self_avg}
\lim_{N\to \infty} g_N(y;\theta) = \lim_{N \to \infty} G^{(x)}_N(y;\theta) = G^{(x)}(y;\theta) \;.
\end{align}
Taking this large $N$ limit in Eq. (\ref{Burgersfinite}), the first term in the rhs drops out and, using (\ref{self_avg}), we arrive at a very simple equation for the Green's function
\begin{align}
\label{BurgersX}
\partial_\theta G^{(x)} + G^{(x)} \partial_y G^{(x)} = 0 \;.
\end{align}
Since $\theta$ is a time-like variable, this equation has to be solved subject to the initial condition 
\begin{eqnarray}\label{CI}
G^{(x)}(y,0) = G^{(x)}_0(y) \;,
\end{eqnarray}
where $G^{(x)}_0(y)$ depends on the choice of the initial positions $\vec{a}$. This is the well known Burgers' equation in fluid dynamics (in the inviscid limit) but in the complex $y$-plane, i.e., a first--order PDE with a non--linear term. 

Note that the Burgers' equation \eqref{BurgersX} that we derived here is in the $(x,\theta)$ coordinates and is already tailored for the bridge configuration because
both coordinates $(x, \theta)$ as defined in Eq. \ref{transformation} already contains the information about the final time $t_f$ and the final flat configuration. In fact,
we recall that this Burgers' equation has been derived starting from the effective Langevin equation (\ref{Langevinx}), which holds only for the bridge configuration, with a
flat final density. On the other hand, for Dyson Brownian motion evolving via Eq. (\ref{model}) one can define a similar Green's function
\bea \label{G_DBM}
G_{\rm DBM}(z;t) = \frac{1}{N} \sum_{i=1}^N \frac{1}{z-\lambda_i(t)} \;.
\eea 
It is then easy to show that, in the large $N$ limit, this Green's function also satisfies the inviscid~Burgers' equation~\cite{BN2010:SHOCKRMT,allez2012invariant,BGNW2015:HERMDETS,krajenbrink2020tilted}
\bea \label{Burger_DBM}
\partial_t G_{\rm DBM}(z;t) + G_{\rm DBM}(z;t) \; \partial_z G_{\rm DBM}(z;t) = 0 \;,
\eea
starting from some initial configuration $G_{\rm DBM}(z;0)$. However, this Green's function does not hold for the bridge since it has no information on the future, in particular 
on the condition at $t=t_f$. Thus, this well known canonical Burgers'~equation has nothing to do with the Burgers' equation for the vicious bridge (\ref{BurgersX}) that we derived above. 
Let us emphasize once more that a crucial ingredient leading to the derivation of this Burgers' equation for the bridge is the use of the effective Langevin equation (\ref{Langevinlambda})
that automatically took the bridge constraint into account. Furthermore, this bridge Burgers' equation (\ref{BurgersX}) holds in the transformed coordinates $(x, \theta)$ and not in the original
coordinates~$(\lambda, t)$.

\subsection{Solution of the Burgers' equation}

\begin{figure}[t]
\includegraphics[width = 0.5\linewidth]{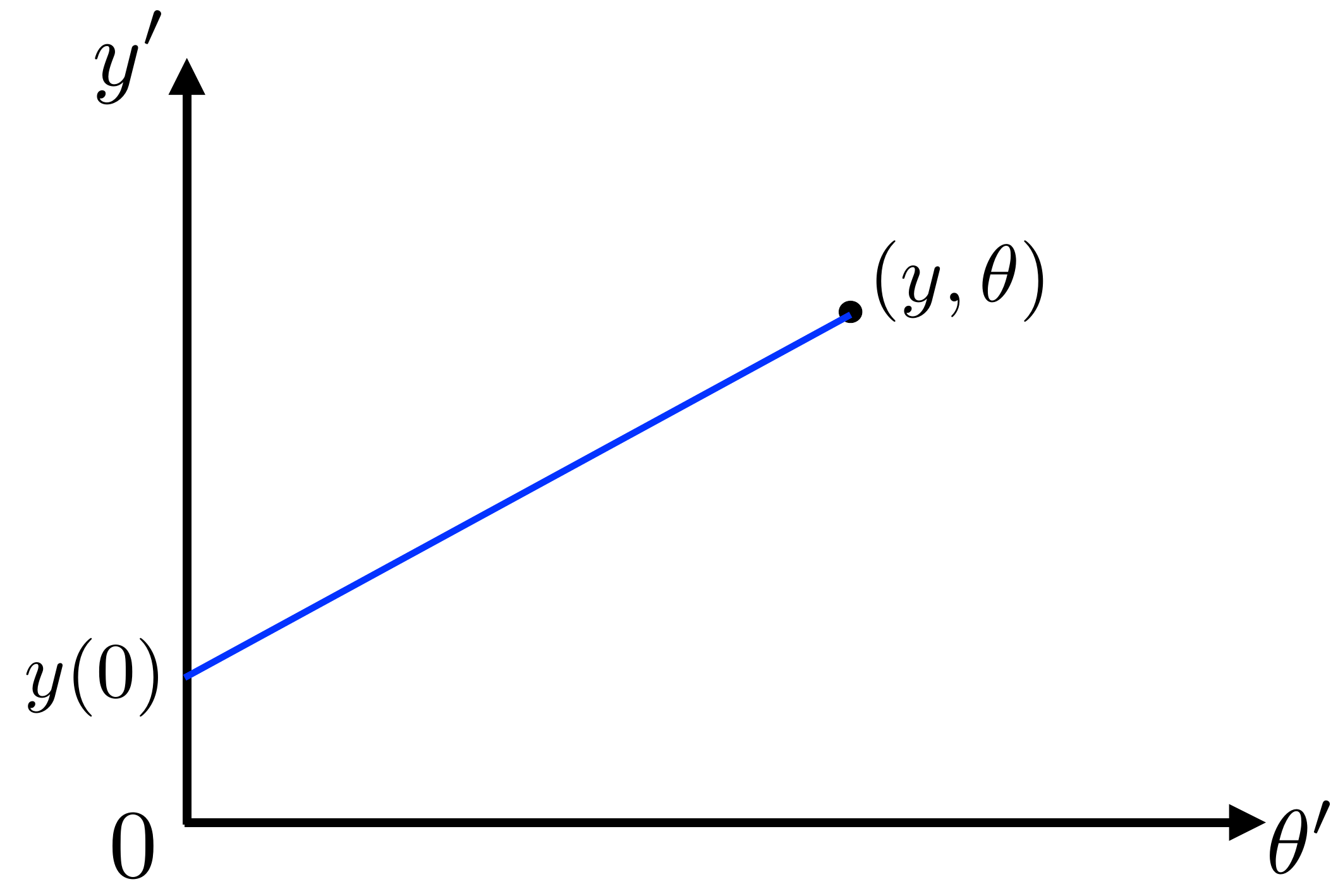}
\caption{Illustration of the characteristic (\ref{char2}) in the $(y',\theta')$ plane.}\label{Fig:char}
\end{figure}

The Burgers' equation \eqref{BurgersX}  can be solved by the standard method of characteristics. Consider the $(y',\theta')$ plane with $(y,\theta)$ being one particular point on this plane. 
We consider a curve $y(\theta')$ in this plane that passes though the point $(y,\theta)$ at time $\theta'=\theta$ (see Fig. \ref{Fig:char}). We choose this curve such that $G^{(x)}(y(\theta');\theta')$ is a constant 
along this curve. Therefore
\begin{eqnarray}\label{char1}
\frac{d}{d\theta'}\,G^{(x)}(y(\theta');\theta') = \frac{\partial G^{(x)}}{\partial \theta'} + \frac{\partial G^{(x)}}{\partial y} \frac{dy(\theta')}{d\theta'}  \;.
\end{eqnarray}
Since $G^{(x)}$ is chosen to be a constant along this curve $y(\theta')$, this total derivative (\ref{char1}) is zero. By comparing the rhs of (\ref{char1}) with 
Eq. (\ref{BurgersX}), we see that curve $y(\theta')$ (called the ``characteristics'') must satisfy the equation of motion
\begin{eqnarray} \label{char_eom}
\frac{dy(\theta')}{d\theta'} = G^{(x)}(y;\theta) \;.
\end{eqnarray} 
This equation is trivial to solve since $G^{(x)}(y;\theta)$ is a constant along the curve and we get a linear characteristic
\begin{eqnarray}\label{char2}
y(\theta') = y(0) + \theta' \, G^{(x)}(y;\theta) \;.
\end{eqnarray}
Along this characteristics we then have
\begin{eqnarray}\label{char3}
G^{(x)}(y,\theta) = G^{(x)}(y(\theta'),\theta') \;, \; \forall \; 0 \leq \theta' \leq \theta \;.
\end{eqnarray}
Setting $\theta'=0$ in Eq. (\ref{char3}) we get 
\begin{eqnarray}\label{char4}
G^{(x)}(y,\theta) = G^{(x)}(y(0),0) = G^{(x)}_0(y(0)) \;,
\end{eqnarray}
where $G^{(x)}_0$ is the initial Green's function (\ref{CI}). Furthermore, setting $\theta' = \theta$ in Eq. (\ref{char2}) and using $y(\theta) =y$, we can determine $y(0)$ in terms of the final value 
\begin{eqnarray}\label{char5}
y(0) = y - \theta \, G^{(x)}(y;\theta) \;.
\end{eqnarray}
Finally substituting this expression for $y(0)$ in Eq. (\ref{char4}) we get an exact self-consistent equation for $G^{(x)}(y;\theta)$
\begin{eqnarray}\label{char_self}
G^{(x)}(y,\theta) = G^{(x)}_0(y - \theta \, G^{(x)}(y;\theta)) 
\end{eqnarray}
where $G^{(x)}_0(\xi)$ is a known function, set by the initial condition. Note that the solution is parametric in the following sense. We can rewrite Eq. (\ref{char_self})
in the following way
\begin{eqnarray}\label{char_self_xi}
G^{(x)}(y,\theta) = G_0^{(x)}(\xi) \quad {\rm where} \quad \xi = y(0) = y - \theta G^{(x)}(y,\theta) \;.
\end{eqnarray}
The second equation gives
\bea \label{y_vs_xi}
y = \xi + \theta \, G^{(x)}(y,\theta) = \xi + \theta \, G_0^{(x)}(\xi) \;,
\eea
where in the last equality we have used the first equality in Eq. (\ref{char_self_xi}). Thus, given $y$ and $\theta$, we have to first solve for $\xi$ from Eq. (\ref{y_vs_xi})  and
then substitute this value of $\xi$ to evaluate $G_0^{(x)}(\xi)$ and then (\ref{char_self_xi}) gives us $G^{(x)}(y,\theta)$. This is the parametric recipe to solve for 
the Green's function $G^{(x)}(y,\theta)$. The function $G^{(x)}_0(\xi)$ that corresponds to the initial Green's function plays a central role in the solution. For our problem, this
is given by 
%
\begin{align}
\label{G3init}
G^{(x)}_0(\xi) =  \frac{e^\xi}{t_f} \lim_{N\to \infty} \frac{1}{N} \sum_{i=1}^N \frac{1}{e^\xi - x_i^{(0)}}.
\end{align}
The initial values $x_i^{(0)}$ and $\lambda_i^{(0)} = a_i$ are related through the transformation \eqref{transformation} as $x_i^{(0)} = e^{a_i/t_f}$. In the following section we consider a flat-to-flat VBB where both the starting points and end points have the same flat density, i.e., $a_i = b_i = (i-1)/N$.

\section{Particle density for a flat-to-flat VBB}
\label{sec3}

In this section, we compute the average density both in the $(\lambda,t)$-coordinates (\ref{def_density_lambda}) 
and equivalently in the $(x,\theta)$ coordinates (\ref{rhox}). We consider the $\beta=2$ DBB, that starts and ends
at the flat configuration $a_i = b_i = (i-1)/N$. At $t=0$, and in the large $N$ limit, the initial condition thus
corresponds to a uniform flat density in the $\lambda$-variable
\bea \label{rho_lamb_t0}
\rho^{(\lambda)}(\lambda;0) = 1 \quad {\rm for} \quad 0 \leq \lambda \leq 1 \;,
\eea
and the density vanishes outside the support $\lambda \in [0,1]$. By virtue of the transformation \eqref{change} we have, at $t=0$, $x_i = e^{\lambda_i/t_f}$. Consequently, the density in the $x$-variable at $\theta = 0$ (corresponding to $t=0$) reads
\bea \label{rho_x_t0}
\rho^{(x)}(x;0) = \frac{t_f}{x} \quad {\rm for} \quad 1 \leq x \leq e^{1/t_f} \;,
\eea 
and the density vanishes outside the support $x \in [1, e^{1/t_f}]$. It is then easily to check that the density is normalized in the $x$-variables, i.e., $\int_1^{e^{1/t_f}} \rho^{(x)}(x;0)\, dx = 1$. 

In the subsection A, we show that the equation for the Green's function (\ref{char_self}) and (\ref{G3init}) in the $(\vec{x},\theta)$ coordinates can be written in a compact form. It turns out that it allows an explicit solution only for three values of $\theta = 1, 2$ and $3$, corresponding respectively to $t=t_f/2$, $t=2t_f/3$ and $t=3t_f/4$.  
In subsection B, we first derive explicitly the average density for $\theta = 1$. Subsequently, in subsection C, the explicit results for the average density are derived respectively for $t=2t_f/3$ and $t=3t_f/4$ (and thus equivalently for $t=t_f/3$ and $t=t_f/4$ thanks to the symmetry around $t=t_f/2$).  In subsection D, we compute explicitly how the edges of the support of the average density $[\lambda_-(t), \lambda_+(t)]$ evolve with time $t$ for all $0\leq t \leq t_f$. Finally, in subsection E, we provide some exact results for the moments of the average density for any $0\leq t \leq t_f$.

\subsection{Solution in terms of e-Green's function $H = e^{-G^{(x)}}$}

Formula \eqref{char_self} is the solution to Burgers' equation \eqref{BurgersX} depending only on the initial function \eqref{G3init}. In the flat case $a_i = \frac{1}{N}(i-1)$ it is found by the Euler-Maclaurin formula that
\begin{align}
\label{initcond}
G^{(x)}(y;0)=G^{(x)}_{0}(y) = \lim_{N\to \infty} \frac{e^y}{N t_f} \sum_{i=1}^N \frac{1}{e^y - e^{\frac{i-1}{Nt_f}}} = \frac{e^y}{t_f} \int_0^1 du \frac{1}{e^y - e^{u/t_f} } =  \ln \left ( \frac{1- e^y}{1-e^{y-1/t_f} } \right ) \;,
\end{align}
which is defined everywhere in the complex $y$-plane, with a cut on the real $y$-axis over the interval 
$y \in (0,1/t_f)$. We plug the initial function \eqref{initcond} into the solution \eqref{char_self} and obtain 
\begin{eqnarray}\label{self_flat}
e^{G^{(x)}} \left ( 1 - e^{y-\theta G^{(x)}-1/t_f} \right ) = 1 - e^{y-\theta G^{(x)}} \;.
\end{eqnarray}
This form (\ref{self_flat}) leads us naturally to define an exponentiated Green's (or e-Green's) function 
\begin{eqnarray}\label{def_H}
H(w;\theta) = e^{-G^{(x)}(y;\theta)} \quad, \quad {\rm where} \quad w = e^y \quad {\rm and} \quad T = e^{-1/t_f} \;.
\end{eqnarray}
This then gives us a compact transcendental equation for $H(w;\theta)$
\begin{align}
\label{Hsol2}
H^\theta = \frac{1}{w} \frac{1-H}{T-H} \;,
\end{align}
where we recall that $\theta = t/(t_f-t)$.

Although for general time $\theta$, the e-Green's function $H$ cannot be found by analytic methods, we will still extract useful information out of it. In particular, the Sochocki-Plemejl formula giving the particle density $\rho^{(x)}$ in terms of the e-Green's function $\rho^{(x)} (x;\theta) = -\frac{t_f}{\pi} \lim_{\epsilon \to 0_+} \text{Im} \left [  \frac{1}{w} \ln H(w;\theta) \right ]_{w = x-i\epsilon}$ is found as (see Appendix \ref{appC})
\begin{align}
\label{rhoform}
\rho^{(x)} (x;\theta) = -\frac{t_f}{\pi x} \lim_{\epsilon \to 0_+} \arg H(x-i\epsilon;\theta) \;.
\end{align}
The scheme for obtaining particles densities in both $(\lambda,t)$ and $(x,\theta)$ variables is the following -- compute the $H$ by solving Eq. \eqref{Hsol2}, use Sochocki-Plemejl formula \eqref{rhoform} to find $\rho^{(x)}$ and lastly compute $\rho^{(\lambda)}$ by formula \eqref{rel}. 

\subsection{Particle density for $t=t_f/2$ ($\theta=1$)}
\label{sectheta1}
Since both initial and final positions are equal, the instant $t=t_f/2$ is special due to the symmetry $t \to t_f-t$ which makes the calculation of the density easier. When $t=t_f/2$, which corresponds to $\theta=t/(t_f-t)=1$, Eq. (\ref{Hsol2}) reduces to a  quadratic equation
\begin{align*}
 wH^2 - wTH - H + 1 = 0,
\end{align*}
which has two explicit solutions $H_\pm(w;\theta=1) = \frac{1}{2w} \left ( 1 + wT \pm T\sqrt{-r_1(w)} \right )$  where 
\begin{eqnarray}\label{def_r1}
r_1(w) = (w^{(1)}_+-w)(w-w^{(1)}_-) \quad , \quad  {\rm with} \quad w^{(1)}_\pm = \frac{2-T \pm 2\sqrt{1-T}}{T^2} \;.
\end{eqnarray}
To choose the correct root, we proceed as follows. From Eq. (\ref{greenx}), we see that, as $y \to \infty$, $G^{(x)}(y;\theta=1) \to 1/t_f$. Consequently $H(w=e^y;\theta) = e^{-G^{(x)}(y;\theta)} \to e^{-1/t_f} = T$. Hence, as $w \to \infty$, we must have $H(w;\theta) \to T$. Therefore we choose $H(w;\theta=1) =H_+(w;\theta=1)$. 

\begin{figure}[h]
\includegraphics[scale=.4]{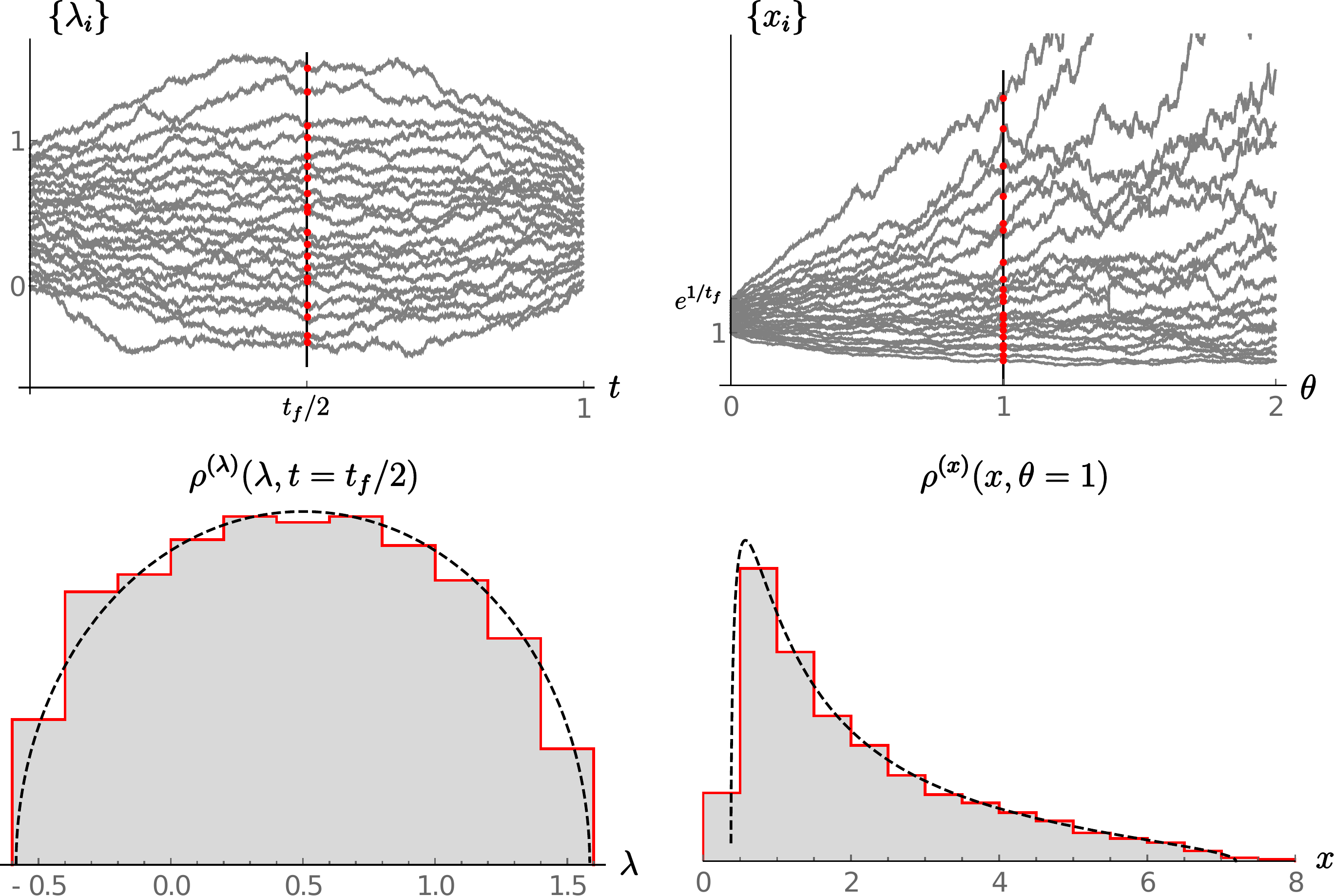}
\caption{Plots of sample trajectories and position densities in $(\lambda,t)$ (left column) and $(x,\theta)$ (right column) variables for a flat-to-flat Brownian bridge. In the plots on the top we present sample trajectories as gray lines with black vertical line denoting the probing times $t = t_f/2$ and $\theta=1$. For these times, we draw both analytic (dashed black line) and numeric (solid red line histograms) position densities given by formulas \eqref{rhotheta1} and \eqref{rholambda}. Simulations were made for $t_f =1$ and $N=20$ by integrating the effective Langevin equations (\ref{Langevinlambda}) and (\ref{Langevinx}).}
\label{fig1}
\end{figure}

To use the formula for the density in \eqref{rhoform}, we set $w=x-i\epsilon$ and expand around $\epsilon=0$, yielding
\begin{align}
\label{Hexp}
H_+ = C_1 + i C_2 \epsilon + \mathcal{O}(\epsilon^2),
\end{align}
where $C_1 = H_+(x) = \frac{1 + xT + T\sqrt{-r_1(x)}}{2x} $, $C_2 = -H_+'(x) = \frac{1-2x+Tx+T\sqrt{-r_1(x)}}{2x^2T\sqrt{-r_1(x)}}$. We consider two cases:
\begin{itemize}
\item For $r_1(x)<0$ or when $x \notin (w^{(1)}_-,w^{(1)}_+ )$, both $C_1$ and $C_2$ are real functions and so  the following result holds
\begin{align}
\label{case1}
\lim_{\epsilon\to 0_+} \arg H(x-i\epsilon;\theta) = \lim_{\epsilon\to 0_+} \arctan \left [ \frac{C_2}{C_1}\epsilon \right ] = 0.
\end{align}
Consequently, the density vanishes for arguments outside of the interval $(w^{(1)}_-,w^{(1)}_+ )$.
\item For $r_1(x)>0$ or when $x \in (w^{(1)}_-,w^{(1)}_+ )$, the square--root term $\sqrt{-r_1(x)}$ becomes imaginary $\sqrt{-r_1(x)} = \sigma i \sqrt{r_1(x)}$ where $\sigma = \pm 1$ is the square--root branch parameter and will be fixed later. Now the expansion \eqref{Hexp} is no longer properly ordered into real and imaginary parts. Instead, we have $C_1 = \frac{1+xT}{2x} + i \sigma T \frac{\sqrt{r_1(x)}}{2x}, C_2 = \frac{1}{2x^2} - \sigma i \frac{1-2x+Tx}{2x^2 T\sqrt{r_1(x)}}$ which renders $H_+ = C_1' + i C_2' + \mathcal{O}(\epsilon^2)$ with $C_1' = \frac{1+xT}{2x} + \sigma \epsilon \frac{1-2x+Tx}{2x^2T\sqrt{r_1(x)}}$, $C_2' = \sigma T\frac{\sqrt{r_1(x)}}{2x} + \frac{\epsilon}{2x^2}$. This rearrangement makes the argument of $H_+$ non--zero inside the interval $(w^{(1)}_-,w^{(1)}_+ )$:
\begin{align}
\label{case2}
\lim_{\epsilon\to 0_+} \arg H_+ = \lim_{\epsilon\to 0_+} \arctan \left [ \frac{C_2'}{C_1'} \right ] = \arctan \left [ \frac{\sigma T\sqrt{r_1(x)}}{1+xT}\right ].
\end{align}
We fix $\sigma = -1$ by demanding that the argument (and the overall density) be positive.
\end{itemize}
We combine Eqs. \eqref{case1} with \eqref{case2} resulting in a particle density bounded to a finite interval:
\begin{align}
\label{rhotheta1}
\rho^{(x)}(x;\theta=1) = \begin{cases} 0 &, \quad  x \notin (w^{(1)}_-,w^{(1)}_+ ) \\ \frac{t_f}{\pi x} \arctan \left [ \frac{T\sqrt{r_1(x)}}{1+xT}\right ]  &, \quad x \in (w^{(1)}_-,w^{(1)}_+ )\;.  \end{cases}
\end{align}
The density is continuous as the function vanishes at the boundaries $r_1(w_\pm^{(1)}) = 0$ and so does $\arctan(0)=0$. The density in $\lambda$ variable is in turn given by Eq. \eqref{rel} as:
\begin{align}
\label{rholambda}
\rho^{(\lambda)}\left (\lambda;t=t_f/2 \right ) = \begin{cases} 0 &, \quad \lambda \notin \left (\lambda^{(1)}_-,\lambda^{(1)}_+ \right ) \\ \frac{2}{\pi} \arctan \left ( T \frac{\sqrt{r_1\left (e^{2\lambda/t_f} \right )}}{1+Te^{2\lambda/t_f}}\right ) &, \quad  \lambda \in \left ( \lambda^{(1)}_- , \lambda^{(1)}_+ \right )\;, \end{cases} 
\end{align}
where the endpoints are given by $\lambda^{(1)}_\pm = \frac{1}{2} \pm \frac{t_f}{2} \text{arccosh}\left (\frac{2-T}{T} \right )$ and are symmetric around $\lambda = 1/2$. We recall that, in formula (\ref{rholambda}) the function $r_1(w)$ is given in Eq. (\ref{def_r1}) and $T = e^{-1/t_f}$. In Fig. \ref{fig1} we plot both densities along with sample trajectories obtained by solving numerically the effective Langevin equations (\ref{Langevinlambda}) and (\ref{Langevinx}). 

As already commented in the introduction, this solution for the special case $t=t_f/2$ appeared before in the literature in different contexts, including Chern-Simons theory and matrix models  \cite{marino2005chern,Mar2006:MATRIXMODELSSTRINGS} as well as in the theory of Stieltjes--Wigert orthogonal polynomials \cite{forrester2020global}, where this result in Eq. (\ref{rholambda}) was derived by different methods.  

Given the explicit expressions for the average densities in (\ref{rhotheta1}) and (\ref{rholambda}), one can also compute the moments of the density. Indeed, in the case of the Gaussian ensembles of Random Matrix Theory, the moments of the Wigner semi-circle density have been computed explicitly in terms of Catalan numbers. We show in Appendix \ref{appD} that the Catalan numbers also appear for this density, albeit in a more complicated way. 


\subsection{Particle densities for $t=\frac{2}{3} t_f$ and $t= \frac{3}{4} t_f$ ($\theta=2$ and $\theta=3$)}
\label{sectheta23}
\begin{figure}[h]
\includegraphics[scale=.4]{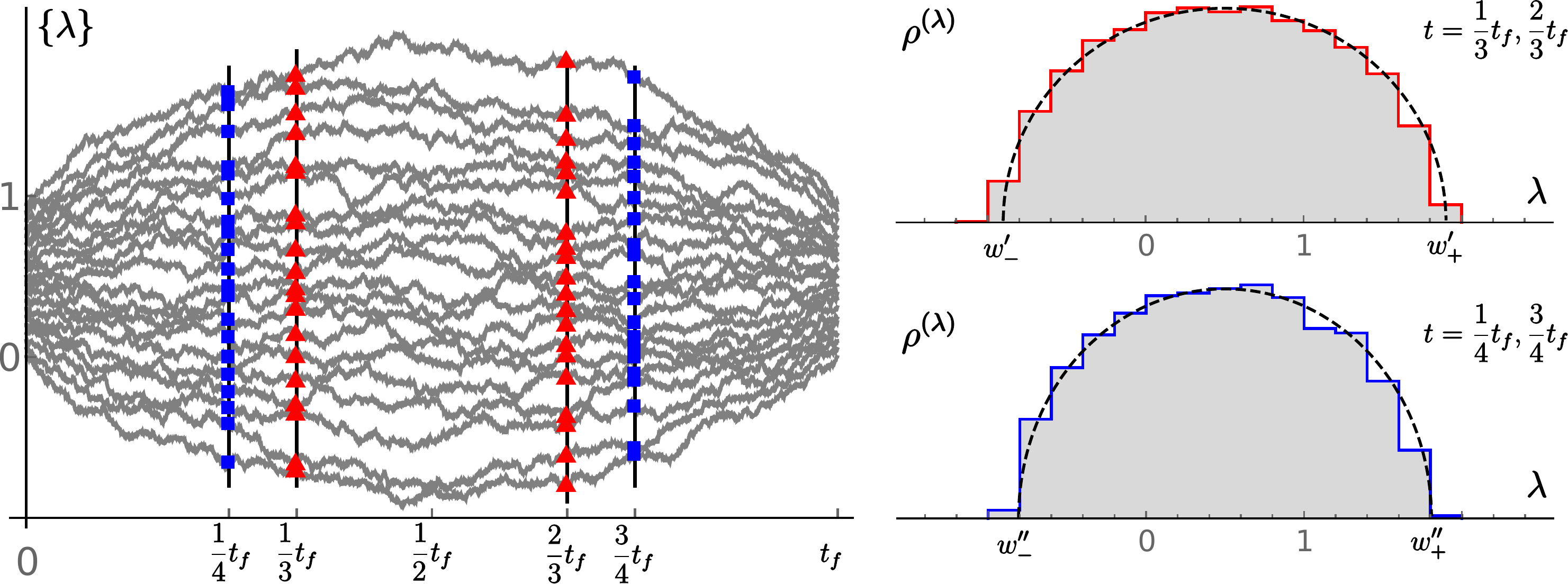}
\caption{The plot on the left shows sample trajectories as gray lines with red triangles denoting the positions at times $t = \frac{2}{3} t_f, \frac{1}{3} t_f$ and blue circles at times $t = \frac{3}{4} t_f, \frac{1}{4} t_f$. On the right we present both analytic (dashed black lines) and numeric (solid red line for $t = \frac{2}{3} t_f, \frac{1}{3} t_f$ and solid blue line for $t = \frac{3}{4} t_f, \frac{1}{4} t_f$) particle densities $\rho^{(\lambda)}$ given by formulas \eqref{rholambda2} and \eqref{rholambda3}. Simulations were made for $t_f = 2$ and $N=20$ by integrating the effective Langevin equations (\ref{Langevinlambda}) and (\ref{Langevinx}).}
\label{fig2}
\end{figure}

We turn to the study of other special cases where the solution to Eq. \eqref{Hsol2} is possible. We find indeed two special values $\theta=2$ and $\theta=3$ (corresponding to $t=\frac{2}{3} t_f$ and $t = \frac{3}{4} t_f$ respectively) where the equation for $H$ \eqref{Hsol2} is cubic and quartic respectively and thus amenable to explicit solutions. In Fig. \ref{fig2} we plot the densities for both times.
\paragraph{Average density for $t=2\,t_f/3$.} Using $\theta = t/(t_f-t)$, we see that this case corresponds to $\theta = 2$. Here, Eq.~\eqref{Hsol2} for the e-Green's function is cubic and reads
\begin{align}
w H^3 -wTH^2 -H +1 = 0 \;. \label{eqtheta2}
\end{align}
Three solutions of Eq. \eqref{eqtheta2} are found by Cardano formulas \cite{enwiki:1005257159} and the one with correct $w \to \infty$ behaviour $H \sim T$ is picked out:
\begin{align}
\label{Hexpansion}
H = \frac{T}{3} + \frac{e^{-i\pi/3}(3+T^2 w)}{3R_2(w)^{1/3}} + \frac{e^{i\pi/3}R_2(w)^{1/3}}{3w},
\end{align}
where $R_2(w) = (3Tw)^{3/2}\sqrt{r_2(w)} - (T w)^3 + \frac{9}{2} (3-T)w^2$ and $r_2(w) = (w_+^{(2)} - w)(w-w_-^{(2)})$. The endpoints are given by $w_{\pm}^{(2)} = \frac{\left (3-T\pm \delta_2(T) \right ) \left (3-3T\pm \delta_2(T) \right )^2}{2T^3(1-T\pm \delta_2(T))^2}$ with $\delta_2(T) = \sqrt{(9-T)(1-T)}$. We consider two cases depending on the endpoints $w_\pm^{(2)}$:
\begin{itemize}
\item For $x\notin (w_-^{(2)} , w_+^{(2)})$, the binomial $r_2(x)<0$ and so the function $R_2(x)$ becomes complex. Despite that, the $H(x)>0$ itself is a purely real and positive function. In that case, $\lim_{\epsilon \to 0_+} \arg H(x-i\epsilon) = 0$ and so the density vanishes $\rho^{(x)}(x;\theta=2) = 0$.
\item For $x \in (w_-^{(2)},w_+^{(2)})$ (or when $r_2(x)>0$) also $R_2(x)>0$ and so the function is expanded
\begin{align}
H(x-i\epsilon) = C_1' + i C_2' + \mathcal{O}(\epsilon^2),
\end{align}
where $C_1' = T/3 + \frac{3+T^2 x}{6R_2(x)^{1/3}} + \frac{R_2(x)^{1/3}}{6x} + \mathcal{O}(\epsilon)$ and $C_2' = - \frac{\sqrt{3}(3+T^2 x)}{6R_2(x)^{1/3}} + \frac{\sqrt{3}R_2(x)^{1/3}}{6x} + \mathcal{O}(\epsilon)$.  The argument of $H$ for $x \in (w_-^{(2)}, w_+^{(2)})$ is given by
\begin{align*}
\lim_{\epsilon \to 0_+} \arg H(x-i\epsilon) = \arctan \left ( \frac{C_2'}{C_1'} \right ).
\end{align*}
\end{itemize}
We use formula \eqref{rhoform} to find the particle density for $\theta=2$:
\begin{align}
\label{rhotheta2}
\rho^{(x)}(x;\theta=2) = \begin{cases} 0 & , \quad x \notin (w_-^{(2)},w_+^{(2)}) \\ \frac{t_f}{\pi x} \arctan \left ( \frac{\sqrt{3}\left [ R_2(x)^{2/3} - x(3+T^2 x) \right ]}{x(3 + T^2 x) + 2Tx R_2(x)^{1/3} + R_2(x)^{2/3}} \right ) & , \quad x \in (w_-^{(2)},w_+^{(2)}) \;. \end{cases}
\end{align}
Finally, the density in the $\lambda$-space is obtained from Eq. \eqref{rel}:
\begin{align}
\label{rholambda2}
\rho^{(\lambda)}\left (\lambda;t = 2t_f/3 \right ) = \begin{cases} 0 & , \quad \lambda \notin (\lambda_-^{(2)},\lambda_+^{(2)})\\
\frac{3}{\pi} \arctan \left ( \frac{\sqrt{3}\left [ R_2(e^{3\lambda/t_f})^{2/3} - e^{3\lambda/t_f}(3+T^2 e^{3\lambda/t_f}) \right ]}{e^{3\lambda/t_f}(3 + T^2 e^{3\lambda/t_f}) + 2Te^{3\lambda/t_f} R_2(e^{3\lambda/t_f})^{1/3} + R_2(e^{3\lambda/t_f})^{2/3}} \right ) & , \quad \lambda \in (\lambda_-^{(2)},\lambda_+^{(2)})\;, \end{cases}
\end{align}
where the endpoints $\lambda^{(2)}_\pm = \frac{1}{2} \pm \frac{t_f}{3} \text{arccosh} \frac{9(3+2T) - T^2}{8T^{3/2}}$ with $T = e^{-1/t_f}$ and the function $R_2(w)$ is defined below Eq. (\ref{Hexpansion}). A plot of the density \eqref{rholambda2} is shown in the top-right plot of Fig. \ref{fig2} and it is in very good agreement with the simulations results from the Langevin equations in \eqref{Langevinlambda}. 

\paragraph{Average density for $t=3\,t_f/4$.} This corresponds to $\theta = 3$, using $\theta = t/(t_f-t)$. In this case, the equation for the e-Green's function is quartic $w H^4 -wH^3 -H +1 = 0$. The solution with correct asymptotic behaviour $H \sim T$ when $w \to \infty$ is given by Ferrari formulas \cite{enwiki:1004178910}
\begin{align}
\label{soltheta3}
H(w) = \frac{T}{4} + \frac{1}{2} \sqrt{S(w)} + \frac{1}{2} \sqrt{ \frac{3T^2}{4} - S(w) + \frac{T^3+8/w}{4\sqrt{S(w)}}},
\end{align}
where $S(w) = \frac{T^2}{4} + \frac{4-T}{3R_3(w)^{1/3}} + \frac{R_3(w)^{1/3}}{w}$, $R_3(w) = \frac{w}{2}\left (1+wT^2 + T^2 \sqrt{-r_3(w)} \right )$ and $r_3(w) = (w^{(3)}_+-w)(w-w^{(3)}_-)$. The endpoints are now equal to $w^{(3)}_\pm = \frac{(2(1-T)\pm\delta_3(T))^3(2-T\pm \delta_3(T))}{T^4(1-T\pm\delta_3(T))^3}$ where $\delta_3(T) = \sqrt{(4-T)(1-T)}$. As before, we consider two cases. 
\begin{itemize}
\item For $x \notin (w_-^{(3)},w_+^{(3)})$ we argue that $H(x)$ is a real function. First, it is evident that the term $\sqrt{-r_3(x)}>0$ is positive which results in both $R_3(x)>0$ and $S(x)>0$. These in turn render positive the functions under the square-roots of Eq. \eqref{soltheta3}. As a consequence, the argument $\lim\limits_{\epsilon \to 0_+} \arg H(x-i\epsilon) = 0$ is zero and the density vanishes:
\begin{align*}
 \rho^{(x)}(x;\theta=3) = 0 \;, \qquad x \notin (w_-^{(3)},w_+^{(3)}).
 \end{align*} 
\item For $x \in (w_-^{(3)},w_+^{(3)})$, the square-root $\sqrt{-r_3(x)}$ becomes imaginary and so $R_3(x)$ is now complex. Still, $S(x)$ remains real and in the end only the square root part of $H(x)$ becomes imaginary which we take into account by setting $\sqrt{ \frac{3T^2}{4} - S(w) + \frac{T^3+8/w}{4\sqrt{S(w)}}} = i\sigma \sqrt{ -\frac{3T^2}{4} + S(w) - \frac{T^3+8/w}{4\sqrt{S(w)}}}$ where $\sigma = \pm 1$ encodes the branch. With this reformulation, the expansion \eqref{soltheta3} reads
\begin{align*}
H(x-i\epsilon) = \frac{T}{4} + \frac{1}{2} \sqrt{S(x)} + \frac{i\sigma}{2} \sqrt{ -\frac{3T^2}{4} + S(x) - \frac{T^3+8/x}{4\sqrt{S(x)}}} + \mathcal{O}(\epsilon)
\end{align*}
and so the argument of the function is simply $\lim\limits_{\epsilon \to 0_+} \arg (H-i\epsilon) = \arctan \left ( \frac{\sigma}{2} \frac{`\sqrt{ S(x) - \frac{3T^2}{4} - \frac{T^3+8/x}{4\sqrt{S(x)}}}}{T/4 + \frac{1}{2} \sqrt{S(x)}}\right )$.
\end{itemize}
We set $\sigma=-1$ to obtain a positive particle density
\begin{align}
\rho^{(x)}(x;\theta=3) & = \begin{cases}  0 & , \quad x \notin (w^{(3)}_-,w^{(3)}_+) \\
\frac{t_f}{\pi x} \arctan \left ( \frac{\sqrt{ S(x)^{3/2} - \frac{3T^2}{4} S(x)^{1/2} - \frac{T^3}{4}-\frac{2}{x}}}{\frac{T}{2} S(x)^{1/4} + S(x)^{3/4} }\right ) & , \quad x \in (w^{(3)}_-,w^{(3)}_+) \;, \end{cases} \label{rhotheta3}
\end{align}
which in the $\lambda$-space is given by
\begin{align}
\label{rholambda3}
\rho^{(\lambda)}\left (\lambda;t = 3t_f/4 \right ) & = \begin{cases} 0 & , \quad \lambda \notin (\lambda_-^{(3)},\lambda_+^{(3)}) \\ \frac{4}{\pi} \arctan \left ( \frac{\sqrt{ S\left (e^{4\lambda/t_f}\right )^{3/2} - \frac{3T^2}{4} S\left (e^{4\lambda/t_f} \right )^{1/2} - \frac{T^3}{4}-\frac{2}{e^{4\lambda/t_f}}}}{\frac{T}{2} S\left ( e^{4\lambda/t_f} \right )^{1/4} + S\left ( e^{4\lambda/t_f} \right )^{3/4} }\right ) & , \quad \lambda \in (\lambda_-^{(3)},\lambda_+^{(3)})\;, \end{cases} 
\end{align}
where $\lambda_\pm^{(3)} = \frac{1}{2} + \frac{t_f}{4}\text{arccosh} \frac{32(4-3T) - T^2(3+2T)}{27T^2}$ with $T = e^{-1/t_f}$ and where the function $S(w)$ is defined below Eq. (\ref{soltheta3}). This formula is plotted in the bottom-right plot of Fig. \ref{fig2} and shows excellent agreement with the simulations of the Langevin equations (\ref{Langevinlambda}).

\subsection{Support of the density as a function of time}

For generic value of the renormalised time $\theta$, Eq. (\ref{Hsol2}) for $H$ is hard to solve to obtain the average density explicitly.
However, one can still extract interesting information valid for any $\theta$. In this subsection, we show how to compute the time evolution
of the support of the average density by analysing the parametric solution (\ref{char_self_xi}) and (\ref{y_vs_xi}) of the original Burgers' equation (\ref{BurgersX}).

The edge of the average density can be extracted by adapting a method originally used in \cite{BN2010:SHOCKRMT,FG2016:HYDROSPECTR} for the DBM with $\beta=2$ in the large $N$ limit, which 
we discuss in detail in Appendix \ref{App:E}. This method can be easily adapted to our problem, the only difference being that we work in $(x,\theta)$ coordinates, as
opposed to the $(\lambda, t)$ coordinates in the case of the $\beta=2$ DBM (see Appendix \ref{App:E}). We recall the parametric solution in Eqs. (\ref{char_self_xi}) and (\ref{y_vs_xi}). 
For fixed $y$ and $\theta$ we need to solve for $\xi$ from the following equation
\bea \label{xi_theta}
y = \xi + \theta \, G_0^{(x)}(\xi) \;.
\eea
Using the explicit expression of $G_0^{(x)}(\xi)$ from Eq. (\ref{initcond}), we get
\bea \label{xi_theta_2}
y = \xi + \theta \, \ln \left(\frac{e^\xi-1}{T\,e^\xi - 1} \right) \;,
\eea
where we used $e^{-1/t_f} = T$. From Eqs. (\ref{xi_theta}) and (\ref{xi_theta_2}), we want to extract the edges at  of the support of the average density $x_{\pm}(\theta)$ at ``time'' $\theta$ in the 
$(x,\theta)$-coordinates. Adapting the method outlined in Appendix \ref{App:E} for the $\beta=2$ DBM, we can extract the edges of the support as follows.

In the complex $\xi$-plane, the function $y(\xi)$ in Eq. (\ref{xi_theta_2}) has a cut along the real axis on the interval $\xi \in [0,1/t_f]$. On this interval,
$y(\xi)$ becomes purely imaginary, while it is real everywhere on the real $\xi$-axis outside this cut. In order to analyse the edges of the support of the density, we need to 
analyse the solution in the two intervals where $y(\xi)$ is real, namely $\xi \in (-\infty,0]$ and $\xi \in [1/t_f, +\infty)$. The first interval $\xi \in (-\infty,0]$ will give us information about the lower edge of the
support of the average density, while the second interval $\xi \in [1/t_f, +\infty)$ provides information about the upper edge of the average density. For simplicity, we now focus on the 
second interval $\xi \in [1/t_f,+\infty)$ and derive the result for the upper edge $x_+(\theta)$ (in the $(x,\theta)$-variables) and equivalently for $\lambda_+(t)$ (in the $(\lambda,t)$-variables).

\begin{figure}[t]
\includegraphics[width = 0.5\linewidth]{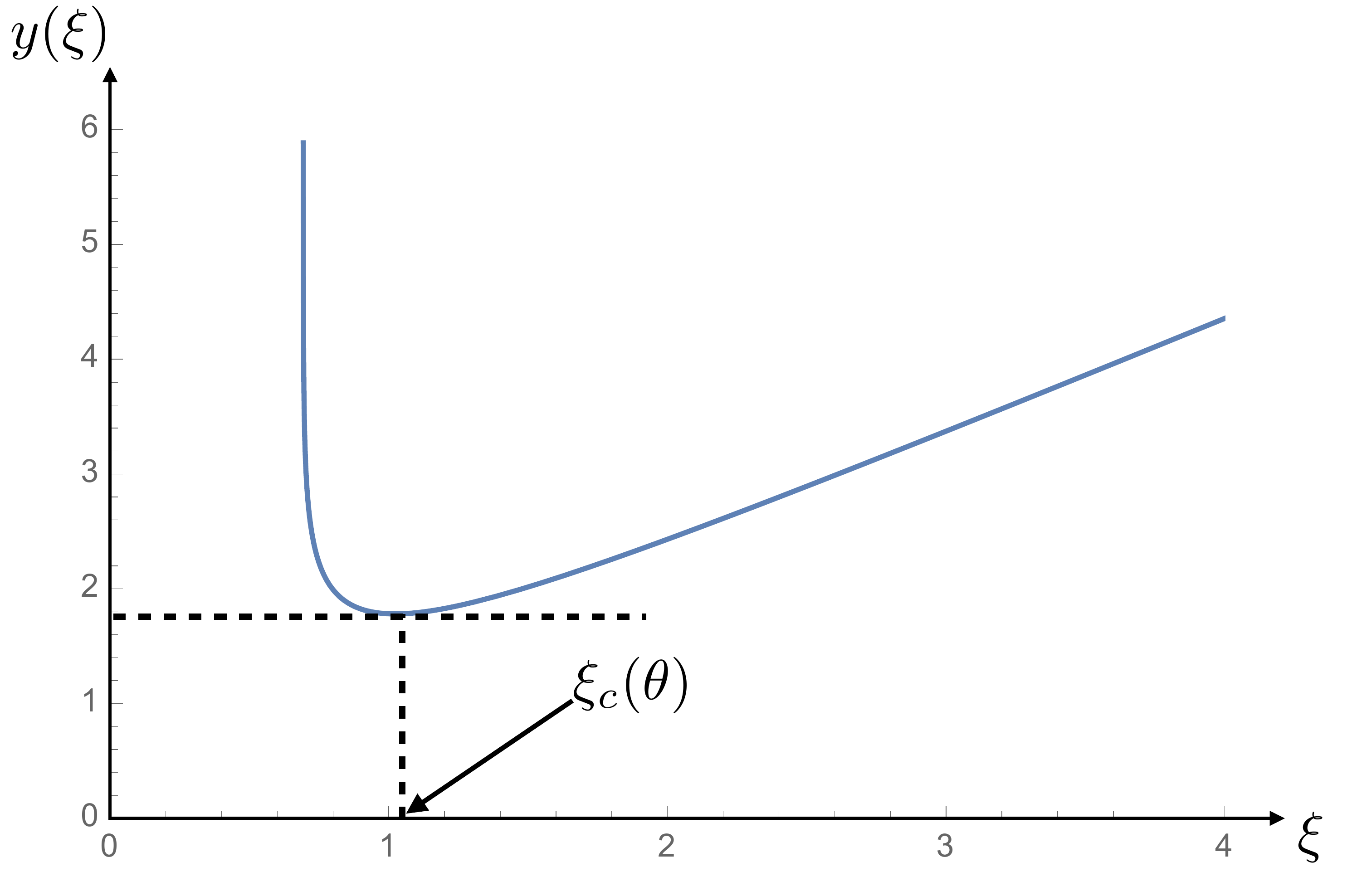}
\caption{Plot of $y(\xi)$ vs $\xi$, as given in Eq. (\ref{xi_theta_2}), for $T=1/2$ and $\theta = 1/2$. In this case $\xi_c(\theta)  = \ln{\left((7+\sqrt{17})/4\right)} = 1.02273\ldots$, where the curve has its minimum.}\label{fig_char_text}
\end{figure}

The function $y(\xi)$, over the interval $\xi \in [1/t_f,+\infty)$, is plotted in Fig. \ref{fig_char_text}. It diverges at the lower edge $\xi_{\rm edge} = 1/t_f$ and increases linearly as $\xi \to \infty$. It clearly has a minimum at $\xi_c(\theta)$ which is determined by setting $dy/d\xi = 0$. This gives
\bea \label{xi_c_theta}
1 + \theta \left(\frac{u}{u-1} - \frac{u\,T}{u\,T - 1} \right) = 0 \quad {\rm where} \quad u = e^{\xi_c(\theta)} \;.
\eea
This is a quadratic equation whose solutions are given by
\bea \label{upm}
u_{\pm}(\theta) =  \frac{1}{2T}\left (  1+T+\theta(1-T) \pm \sqrt{( 1+T+\theta(1-T))^2 - 4T} \right ) \;.
\eea 
Since $\xi_c(\theta) \in [1/t_f, + \infty)$, and $u=e^{\xi_c(\theta)}$, we must have $u(\theta)>1/T$ where $T = e^{-1/t_f}$. This leads us to choose $u_+(\theta)$ as the correct root.  
We now have to substitute $\xi_c(\theta) = \ln u_{+}(\theta)$ in Eq. (\ref{xi_theta_2}) to obtain the edge $y_c(\theta)$. For this it is convenient to first exponentiate the relation in Eq.~(\ref{xi_theta_2}) and rewrite it as
\bea \label{xi_theta_3}
e^{y_c(\theta)} = u_{+}(\theta) \left( \frac{u_+(\theta)-1}{u_+(\theta)\, T - 1}\right)^\theta \;.
\eea
Using $x_i = e^{y_i}$ then gives the upper boundary of the density in the $(x, \theta)$ coordinates
\bea \label{xplus}
x_+(\theta) = e^{y_c(\theta)} = u_{+}(\theta) \left( \frac{u_+(\theta)-1}{u_+(\theta)\, T - 1}\right)^\theta \;,
\eea
where $u_+(\theta)$ is given in Eq. (\ref{upm}). This concludes the derivation of the upper edge.

\begin{center}
\begin{figure}[t]
\includegraphics[scale=.4]{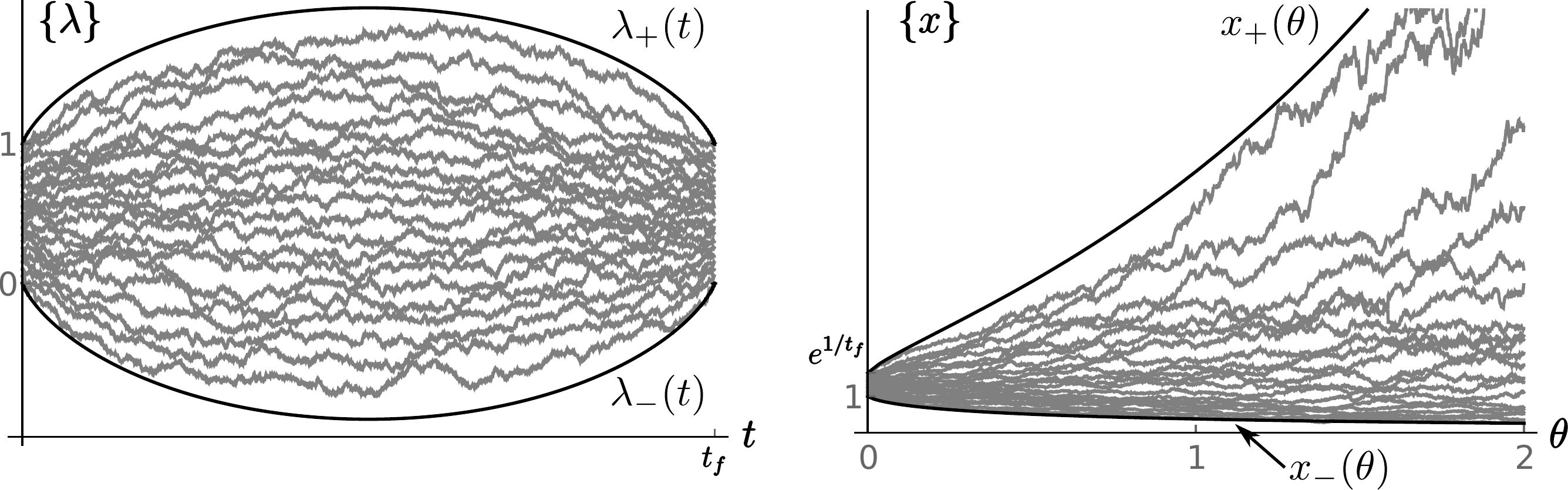}
\caption{Plot of sample trajectories (gray lines) in the $(\lambda,t)$-coordinates (left panel) and in the $(x,\theta)$-coordinates (right panel). The solid lines correspond to the edges of the support of the average density. In the left panel, the lower (respectively upper) solid line corresponds to $\lambda_-(t)$ (respectively $\lambda_+(t)$) given in Eqs. (\ref{endpointslambda}). On the right panel, the lower solid line corresponds to $x_-(\theta)$ in Eq. (\ref{xminus}), while the upper solid line corresponds to $x_+(\theta)$ in Eq. (\ref{xplus}). Simulations were made for $t_f = 2$ and $N=20$ by integrating the effective Langevin equations (\ref{Langevinlambda}) and (\ref{Langevinx}).}
\label{fig3}
\end{figure}
\end{center}

To compute the lower edge of the density, we need to repeat the same reasoning as above except that we
have to consider instead the first interval $\xi \in (-\infty,0]$. In this case, repeating the similar lines of reasoning,
we actually arrive at the same equation (\ref{xi_c_theta}). However, since  
$\xi_c(\theta) \in (-\infty,0]$, we must have $u(\theta) = e^{\xi_c(\theta)} \in [0,1]$ and indeed this is precisely given
by the other root $u_-(\theta)$ of Eq. (\ref{upm}). It is easy to check from \eqref{upm} that $u_-(\theta) \in [0,1]$. This then
gives us the lower edge of the support in the $(x,\theta)$ variables   
\bea \label{xminus}
x_-(\theta) = u_{-}(\theta) \left( \frac{u_-(\theta)-1}{u_-(\theta)\, T - 1}\right)^\theta \;.
\eea
In Fig. \ref{fig3} (right panel), we show a plot of these two boundaries $x_{\pm}(\theta)$ as a function of $\theta$.

Now we can translate these results in terms of the $(\lambda,t)$ coordinates, as defined in Eq. (\ref{transformation}). Hence we set $\lambda_+(t)  = \frac{t_f}{1+\theta} \, y_c(\theta)$ and $t = t_f \frac{\theta}{1+\theta}$ where $y_c(\theta)$ is given in Eq. (\ref{xi_theta_3}). After a few steps of algebra, we can write $\lambda_+(t)$ explicitly. Similarly, for the lower support, we choose $u(\theta) = u_-(\theta)$ in Eq. (\ref{upm}) and replace $u_+(\theta)$ by $u_-(\theta)$ in Eq.~(\ref{xi_theta_3}). Repeating the same manipulations gives $\lambda_-(t)$. Together, they read
\begin{align}
\lambda_\pm(t) = \frac{1}{2} \pm \left [ t_f \text{arccosh} \left ( \frac{1}{\sqrt{T}} \frac{t_f +T(t_f-2t)}{2(t_f-t)} \right ) - t~ \text{arccosh} \left ( \frac{(t_f-t)^2+t^2-T(t_f-2t)^2}{2t(t_f-t)} \right ) \right ] \quad {\rm where} \quad T = e^{-1/t_f} \;.\label{endpointslambda}
\end{align}
These boundaries $\lambda_{\pm}(t)$ are plotted in Fig. \ref{fig3} (left panel). For the special values $\theta=1,2,3$, i.e., $t=t_f/2, t_f/3$ and $t_f/4$ respectively,  
considered in the previous section, we recover the endpoints found in explicit formulae for the average density in Eqs. \eqref{rhotheta1}, \eqref{rhotheta2} and \eqref{rhotheta3}. 

Note that in this paper we have chosen the initial and the final flat configurations to have a unit density supported over the interval $[0,1]$, i.e. $a_i = b_i = (i-1)/N$. 
One can easily generalise our results to the case where $a_i = b_i = \alpha (i-1)/N$ so that the initial and final densities are flat but with value $1/\alpha$ supported
over the interval $[0,\alpha]$. We do not present the details here but we have verified that, taking the $\alpha \to 0$ limit with $t_f$ fixed, we recover the known results 
for a pure ``watermelon'' configuration, where all the particles start at the origin at $t=0$ and end up at the origin at $t=t_f$, i.e., ``point to point'' as opposed to ``flat to flat''
configurations. For example, we have checked that the formula for the boundaries of the support in Eq. \eqref{endpointslambda}, appropriately generalized to $\alpha$,
reduces, to leading order as $\alpha \to 0$, to 
\bea \label{watermelon}
\lambda_{\pm}(t) = \pm \sqrt{\frac{4 \alpha \, t\,(t_f-t)}{t_f}} \;,
\eea
which is exactly the boundary of a point-to-point watermelon with diffusion constant $D = \alpha/(2N)$ (see e.g. \cite{SMCR2008:BROWNMAXIMAL}).

\subsection{Moments of the average density for arbitrary time $0\leq t \leq t_f$}
\label{momentsthetageneral}

Another usefulness of the general result for $H$ in Eq.~\eqref{Hsol2} is that it allows to calculate the moments of the average
density for arbitrary $\theta$, which corresponds to arbitrary real time through the relation $t = \theta \,t_f/(1+\theta)$.  
To compute this, we recall the definition of the Green's function given by Eq. \eqref{greenx} (with $e^z = w$) and expand it in power of $1/w$ 
\begin{align}
\label{Gexp}
G^{(x)}(z;\theta) = \frac{1}{t_f} \sum_{k=0}^\infty \frac{m^{(x)}_{k,\theta}}{w^k},
\end{align}
where $m^{(x)}_{k,\theta} = \int dx x^k \rho^{(x)}(x;\theta)$ are the moments of the density. We plug $H = e^{-G^{(x)}}$ into the solution \eqref{Hsol2} and expand in powers of $1/w$. By matching the powers of $1/w$ on both sides of \eqref{Hsol2}, we obtain a recursion relation between the moments which can then be used to compute the moments successively. We list the first few moments in Table \ref{tab1}. We cross-checked the algorithm up to 10-th order with explicit moment formulas $m^{(x)}_{k,\theta=1}$, for $\theta = 1$, given by Eq. \eqref{catalanmoments}.
\begin{center}
\begin{table}[h!]
\begin{tabular}{|c|c|}
\hline
$k$ & $m^{(x)}_{k,\theta}/t_f$ \\
\hline
0 & $1/t_f$ \\
\hline
1 & $-\left(T-1\right) T^{-(\theta+1) }$ \\
\hline
2 & $\frac{1}{2} (T-1) T^{-2 (\theta +1)} (-2 \theta +(2 \theta -1) T-1)$ \\
\hline
3 & $-\frac{1}{6} (T-1) T^{-3 (\theta +1)} \left(9 \theta ^2+9 \theta +\left(9 \theta ^2-9 \theta +2\right) T^2+\left(2-18 \theta ^2\right) T+2\right)$ \\
\hline
4 & $\!\begin{aligned}[t] \frac{1}{12} (T-1) T^{-4 (\theta +1)} \Big ( & -32 \theta ^3-48 \theta ^2-22 \theta +\left(32 \theta ^3-48 \theta ^2+22 \theta -3\right)
   T^3+\left(-96 \theta ^3+48 \theta ^2+6 \theta -3\right) T^2 \\ & +\left(96 \theta ^3+48 \theta ^2-6 \theta -3\right) T-3\Big ) \end{aligned}$ \\
   \hline
5 & $\!\begin{aligned}[t] -\frac{1}{120} (T-1) T^{-5 (\theta +1)} \Big (& 625 \theta ^4+1250 \theta ^3+875 \theta ^2+250 \theta +\left(625 \theta ^4-1250 \theta ^3+875
   \theta ^2-250 \theta +24\right) T^4 \\ & 
   -4 \left(625 \theta ^4-625 \theta ^3+125 \theta ^2+25 \theta -6\right) T^3+6 \left(625 \theta ^4-125 \theta
   ^2+4\right) T^2 \\ 
   & -4 \left(625 \theta ^4+625 \theta ^3+125 \theta ^2-25 \theta -6\right) T+24\Big ) \end{aligned}$ \\
   \hline
\end{tabular}
\caption{List of the five first moments of position density $\rho^{(x)}$ in the flat-to-flat Brownian bridge scenario valid for any time variable $\theta$ and obtained by iteratively solving Eq. \eqref{Hsol2}.}
\label{tab1}
\end{table}
\end{center}

\section{VBB and its relation to other models}
\label{sec4}
Flat-to-flat VBB as described in the present work is also related to other models. We discuss briefly connections to Chern-Simons matrix model, biorthogonal Stieltjes-Wigert ensemble and Muttalib-Borodin ensemble. 

We recall that the joint distribution of the positions $\vec{\lambda}$ of $N$ particles in the VBB at time $t$ is given in Eq. \eqref{P_vicious_bridge} where $P_{{\rm VBM},D}$ is given by the Karlin-McGregor formula in Eq. \eqref{KMG2}. Setting $D = 1/(2N)$ in Eq. \eqref{P_vicious_bridge} we then get
\begin{align}
\label{Ptildepart1}
& P_{{\rm VBB}, D=\frac{1}{2N}}\left (\vec{\lambda},t | \vec{a},0;\vec{b},t_f \right ) \propto \det_{1\leq i,j \leq N} \left ( e^{-\frac{N}{2t}\left (\lambda_i - a_j \right )^2} \right ) \det_{1\leq i,j \leq N} \left ( e^{-\frac{N}{2(t_f-t)}\left (b_i - \lambda_j \right )^2} \right ),
\end{align}
where we keep track of only the $\lambda$-dependent terms. In the case of flat initial and final positions $a_i = b_i = \frac{i-1}{N}$ we obtain
\begin{align}
\label{jpdfbridge}
P_{{\rm VBB}, D=\frac{1}{2N}}\left (\vec{\lambda},t | \vec{a},0;\vec{b},t_f \right ) \propto \prod_{i=1}^N e^{- \frac{Nt_f}{2t(t_f-t)} \lambda_i^2 + \frac{(N-1)t_f}{2t(t_f-t)}\lambda_i }  \prod_{i>j} \sinh \left ( \frac{\lambda_i - \lambda_j}{2t} \right ) \prod_{i>j} \sinh \left ( \frac{\lambda_i - \lambda_j}{2(t_f-t)} \right ) \;.
\end{align} 
The pdf is not symmetric under the exchange $\lambda_i \to -\lambda_i$ as both initial and final conditions lack this symmetry. On the other hand, time reversal symmetry $t \to t_f - t$ is preserved. Performing the shift $\lambda_i \to \lambda_i - \frac{N-1}{2N}$ and setting $t=t_f/2$, the rhs of (\ref{jpdfbridge}) reduces exactly, up to a global proportionality factor, to the statistical weight factor appearing in the partition function of the Chern-Simons model \cite{marino2005chern,Mar2006:MATRIXMODELSSTRINGS}, and is also related to the theory of 
Stieltjes-Wigert orthogonal polynomials \cite{DT2007:SWANDCS,szabo2010chern,TK2012:NONCOLLSW,borot2016asymptotic}. For other values of $t$, the rhs can be identified with the statistical weight in a generalized Chern-Simons model and also can be related to the theory of bi-orthogonal Stieltjes-Wigert polynomials \cite{DT2007:SWANDCS,szabo2010chern}.

There is yet another way of rewriting the joint distribution $P_{{\rm VBB}, D=\frac{1}{2N}}$ in \eqref{Ptildepart1}. For the flat initial conditions $a_i = b_i = (i-1)/N$, one can
evaluate the determinant as in Eqs. (\ref{equi}) and (\ref{KMG3}). In terms of the variables $x_i = e^{\theta\,\lambda_i/t}$ and $\theta = t/(t_f-t)$, the joint distribution can be expressed as
\begin{align}
\label{jpdfbridge2}
{\cal P}_{{\rm VBB}, D=\frac{1}{2N}}\left (\vec{x}, \theta | \vec{a},0;\vec{b},t_f \right ) \propto \prod_{i=1}^N e^{- \frac{Nt_f}{2\theta} (\log x_i)^2 - \log x_i }  \prod_{i>j} (x_i - x_j) \prod_{i>j} \left ( x_i^{1/\theta} - x_j^{1/\theta} \right ) \;.
\end{align}
Thus in terms of the $(x,\theta)$ variables, the VBB joint distribution is also related to Muttalib-Borodin ensemble \cite{Bor1998:BIORTHOG,Mut1995:BIORTH} with parameter $1/\theta$.

\vspace*{0.5cm}
\noindent{\it Comparison with finite $N$ results.} Establishing relations with other models enables some comparisons. Asymptotic results established in present work are juxtaposed with a closed form of the particle densities $\rho_N^{(x)}$ valid for finite $N$:
\begin{align}
\label{rhoexact}
\rho_N^{(x)}(x;\tau,q) = \frac{1}{qN} w(x/q,q) \sum_{n=0}^{N-1} T_n(x/q,\tau,q) R_n(x/q,\tau,q),
\end{align}
where the weight $w(x,q) = \frac{1}{\sqrt{2\pi|\log q|}} \exp \left ( -\frac{(\log x)^2}{2|\log q|} \right )$ and the polynomials read
\begin{align}
T_n (x,\tau,q) & = (-1)^n \frac{\sqrt{(q;q)_n} q^{(n\tau + 1/2)/2}}{(q^\theta;q^\theta)_n} \sum_{l=0}^n (-1)^l \stirlingii{n}{l}_{q^\tau} q^{\tau l (l(\tau + 1) + 1)/2} x^{\tau l}, \nonumber\\
R_n (x,\tau,q) & = (-1)^n \frac{q^{(n\tau + 1/2)/2}}{\sqrt{(q;q)_n}} \sum_{l=0}^n (-1)^l \stirlingii{n}{l}_{q^\tau} q^{l(l(\tau+1)+(1-\tau)+1)/2} x^l, \label{polys}
\end{align}
where $\stirlingii{x}{y}_q$ and $(q;q)_n$ are the q-extensions of the binomial and the Pochhammer symbol respectively. The formula can be found in Eq. (62) of \cite{TK2012:NONCOLLSW} where only a proper rescaling was introduced to correct the (lack of) symmetry in the joint distribution of the VBB \eqref{jpdfbridge2}.

\begin{figure}[h]
\includegraphics[scale=.4]{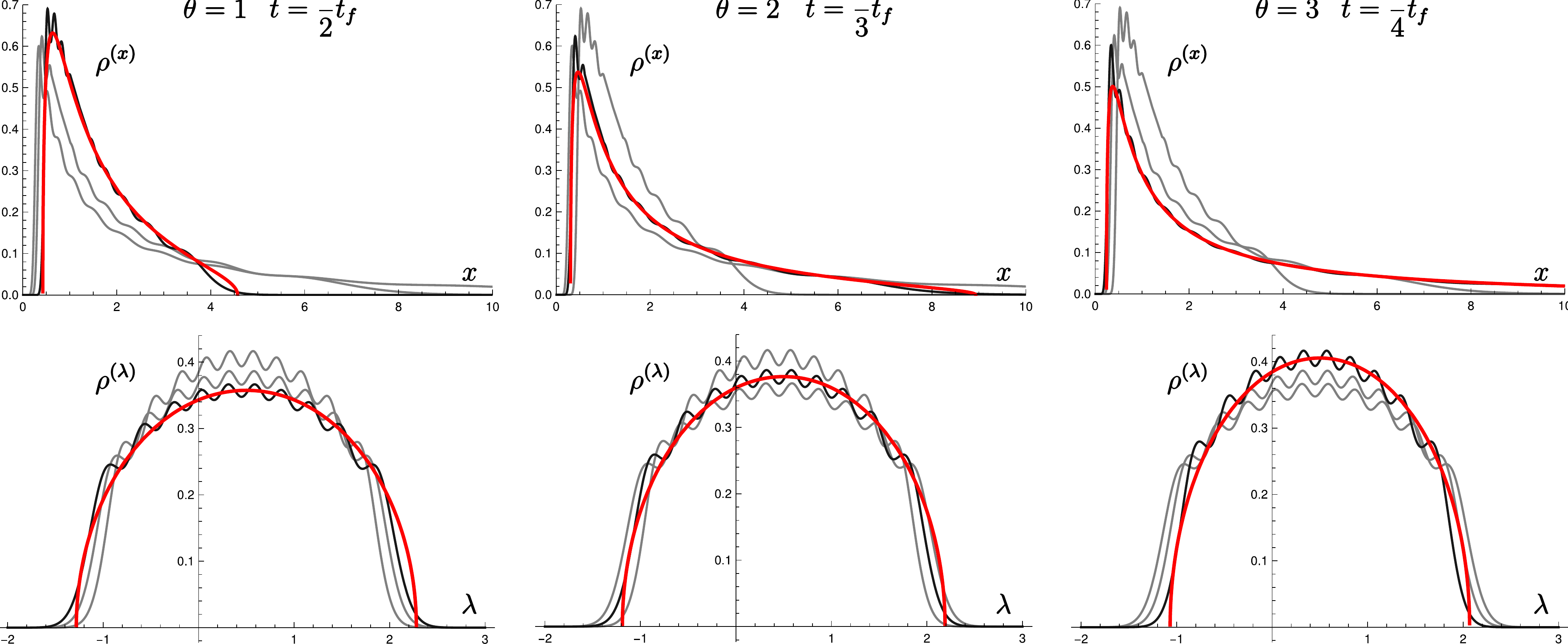}
\caption{Plots of exact position density given by Eq. \eqref{rhoexact} and asymptotic formulas \eqref{rhotheta1},\eqref{rhotheta2} and \eqref{rhotheta3} for times $\theta=1,2,3$ (or $t=\frac{1}{2}t_f,\frac{2}{3}t_f,\frac{3}{4}t_f$). The asymptotic formulas show no oscillatory behaviour, otherwise the matching is very good even for moderate number of particles. The plots were made for $N=10$ and $t_f=3$.}
\label{fig5}
\end{figure}

 In our notation, the $q$ and $\tau$ parameters are equal to $q = e^{- \frac{\theta}{Nt_f}}$ and $\tau = \frac{1}{\theta}$ respectively. The formula in \eqref{rhoexact} is therefore the particle density in the $(x,\theta)$ space and the transformation to $(\lambda,t)$ variables is given by Eq. \eqref{rel}. In Fig. \ref{fig5} we present comparison between asymptotic density formulas valid for times $\theta=1, 2, 3$ found in Eqs.  \eqref{rhotheta1}, \eqref{rhotheta2} and \eqref{rhotheta3} with an exact formula \eqref{rhoexact}.

\vspace*{0.5cm}
\noindent{\it Kernel structure of the underlying determinantal process.} Finally, we end this section by making some remarks on the form of the kernel 
that characterises the determinantal process. The formulae provided below are perhaps a bit formal but they maybe be useful for future large $N$ asymptotic
analysis. A general reader may skip this section, which is intended for experts in the field. 

We start with the joint pdf \eqref{jpdfbridge2} itself
\begin{align*}
P_{\tau,q} \sim \prod_{i=1}^N e^{- \frac{(\log x_i)^2}{2|\log q|}  - \log x_i }  \prod_{i>j} (x_i - x_j) \prod_{i>j} \left ( x_i^{\tau} - x_j^{\tau} \right ),%
\end{align*}
where we reintroduced the parameters $q,\tau$. Our aim is to offer an integral representation of the kernel complementing the density formula \eqref{rhoexact}. To this end, we cast the jpdf into a biorthogonal form
\begin{align*}
P_{\tau,q} \sim \det \left ( \eta_i (x_j) \right )_{i,j=1...N} \det \left ( \xi_i (x_j) \right )_{i,j=1...N},
\end{align*}
where the polynomials are $\eta_i(x) = x^{i-1}$ and $\xi_i(x) = \frac{1}{\sqrt{2\pi |\log q|}}e^{-\frac{(\log x)^2}{2|\log q|}-\log x} x^{\tau(i-1)}$. The following formula found in \cite{BS2006:CHARDETS} gives the kernel function simply related to the average of ratio of determinants
\begin{align}
\label{kernelratio}
\left < \prod_{i=1}^N \frac{v-x_i}{z-x_i} \right > = \int_0^\infty dx \frac{v-x}{z-x} K_N(v,x).
\end{align}
To calculate the ratio we use the following formula given in \cite{FL2016:GINPROD,FGS2018:CHIRALSOURCE}:
\begin{align}
\label{eq0}
& \left < \prod_{i=1}^N \frac{v-x_i}{z-x_i} \right > = \int_0^\infty dx \frac{v-x}{z-x} \sum_{k=1}^N \xi_k(x) \frac{\det g_k(v)}{\det g} , 
\end{align}
where the matrix elements are $g_{ij} = \int_0^\infty dx \eta_i (x) \xi_j (x) = q^{-\frac{1}{2} \left [ i-1 + \tau(j-1) \right ]^2}$ and the matrix $g_k(v)$ is $g$ with $k$-th column replaced by a vector $(1,v,...,v^{N-1})$. Since the elements $g_{ij}$ are quadratic function of the indices, we can rewrite both determinants as Vandermonde terms. To cast the $v$-dependent column into a form compatible with the Vandermonde structure, we use an integral representation of the monomial function $v^{n-1} = \frac{q^{-n^2/2+n(\tau+1)} }{v\sqrt{2\pi |\log q|} } \int_{-\infty}^\infty dw e^{ -\frac{(w+i \log \frac{v}{q^{\tau+1}})^2}{2|\log q|} +inw}$. Using these manipulations, we can rewrite the ratio of determinants as    
\begin{align}
\label{formulaa}
\frac{\det g_k(v)}{\det g} = \frac{1}{v k_k \mu_k \sqrt{2\pi |\log q|}} \int_{-\infty}^\infty dw e^{ -\frac{(w+i \log \frac{v}{q^{\tau+1}})^2}{2|\log q|}+iw} \prod_{j(\neq k)} \frac{\mu_j - e^{iw}}{\mu_j-\mu_k}, 
\end{align}
with $k_j = q^{-\frac{1}{2}(\tau(j-1)-1)^2}$ and $\mu_j = q^{-\tau j}$. In the last step, we extract the $k$-dependent terms from formula \eqref{formulaa} and introduce the sum
\begin{align*}
S = \sum_{k=1}^N \frac{\xi_k(x) }{k_k \mu_k} \prod_{j(\neq k)} \frac{\mu_j - e^{iw}}{\mu_j-\mu_k}.
 \end{align*} 
We plug in the definitions and insert the index $k = -\frac{\log \mu_k}{\tau \log q}$ obtained by inverting the definition of $\mu_k$. This results in a sum which can be expressed through a contour integral
\begin{align}
\label{eq22}
S = - \frac{e^{-\frac{(\log x)^2}{2|\log q|}-\log x}}{\sqrt{2\pi |\log q|}} \oint_C \frac{d u}{2\pi i} x^{-\tau\left (\frac{\log u}{\tau\log q} + 1 \right )} q^{\frac{\left [ \tau \left ( \frac{\log u}{\tau \log q} +1 \right )+1 \right ]^2}{2}} \frac{1}{u(u - e^{iw})} \prod_{j=1}^N \frac{e^{iw}-\mu_j}{u-\mu_j} , 
\end{align}
where the contour $C$ encircles anti-clockwise all poles located at $\mu_j$'s but stays outside both $0$ and $e^{iw}$.
We bring together Eqs. \eqref{eq0} and \eqref{eq22} and use the formula \eqref{kernelratio}:
\begin{align}
\label{kernel}
K_N(x,y) = \frac{1}{2\pi |\log q|x} \oint_C \frac{d u}{2\pi i} \int_{-\infty}^\infty dw e^{ -\frac{\left [w+i \log \left (x/q^{\tau+1} \right ) \right ]^2+\left [\log u - \log \left ( y/q^{\tau+1} \right ) \right ]^2}{2|\log q|}}  \frac{e^{iw}}{u(e^{iw}-u)} \prod_{j=1}^N \frac{e^{iw}-q^{-j\tau}}{u-q^{-j\tau}}.
\end{align}

The particle density is the diagonal part of the kernel. Equivalence with Eq. \eqref{rhoexact} can be established by an identity $\frac{1}{e^{iw}-u)} \prod_{j=1}^N \frac{e^{iw}-q^{-j\tau}}{u-q^{-j\tau}} = \frac{1}{e^{iw}-u} + \sum_{n=0}^{N-1} \frac{\prod_{j=1}^{}(e^{iw}-q^{-j\tau})}{\prod_{j=1}^{n+1} (u-q^{-j\tau})}$. The first term in this identity is vanishing due to the contour integral. Integrals in the remaining sum are decoupled and form integral representations of the polynomials defined by Eqs. \eqref{polys}. The coupled form \eqref{kernel} is however a promising step in conducting asymptotic analysis independently of the Dysonian approach presented in this work. For completeness, the kernels in both $(\lambda,t)$ and $(x,\theta)$ variables are found by setting $q = e^{-\frac{\theta}{Nt_f}}$ and $\tau = 1/\theta$ and using Eq. \eqref{transformation}
\begin{align}
K^{(x)}_N(x,y;\theta) & = \frac{Nt_f}{2\pi \theta x} \int_{-\infty}^\infty dw \oint_C \frac{d u}{2\pi i} e^{ -\frac{Nt_f}{2\theta} \left [ \left (w+i \log x + i \frac{\theta+1}{Nt_f} \right )^2 + \left (\log u - \log y - \frac{\theta+1}{Nt_f}\right )^2 \right ]} \frac{e^{iw}}{u(e^{iw}-u)} \prod_{j=1}^N \frac{e^{iw}-e^{\frac{j}{Nt_f}}}{u-e^{\frac{j}{Nt_f}}},& \\
K^{(\lambda)}_N(\xi,\lambda;t) & = \frac{Nt_f}{2\pi t} \int_{-\infty}^\infty dw \oint_C \frac{d u}{2\pi i} e^{-\frac{Nt_f(t_f-t)}{2t} \left [ \left (w+i\frac{\xi+N^{-1}}{t_f - t} \right )^2 + \left (\log u - \frac{\lambda+N^{-1}}{t_f - t} \right )^2 \right ]} \frac{e^{iw}}{u(e^{iw}-u)} \prod_{j=1}^N \frac{e^{iw}-e^{\frac{j}{Nt_f}}}{u-e^{\frac{j}{Nt_f}}},
\end{align}
where the densities read simply $\rho^{(\lambda)}_N(\lambda;t) = \frac{1}{N} K^{\lambda)}_N(\lambda,\lambda;t) $ and $\rho^{(x)}_N(x;\theta) = \frac{1}{N} K^{(x)}_N(x,x;\theta)$. These kernels form the basic building blocks to compute the correlation functions in the model described by \eqref{jpdfbridge} and \eqref{jpdfbridge2}.

\section{Summary and outlook}

In this paper, we have studied $N$ vicious Brownian bridges propagating from an initial configuration $\vec{a}$ at time $t=0$ to a final configuration $\vec{b}$
at time $t=t_f$, while staying non-intersecting for all $0\leq t \leq t_f$. We mapped this vicious bridge problem exactly to Dyson's Brownian bridges with
Dyson index $\beta=2$ and for the latter we derived an exact effective Langevin equation that generates very efficiently the vicious bridge configurations.  
In particular, for the flat-to-flat configuration in the large $N$ limit, we used this Langevin equation to derive an exact Burgers' equation (in the inviscid limit)   
for the Green's function and provided the solution of this Burgers'~equation at arbitrary time $0\leq t \leq t_f$. We emphasize that this Burgers' equation is 
derived from the effective Langevin equation for the bridge in the transformed $(x,\theta)$ coordinates [see Eq. (\ref{Langevinx})] and already contains
in it the future information about the final bridge condition at $t=t_f$. In this sense, this is different from the canonical Burgers' equation associated to
the $\beta=2$ DBM, which is well known. From this Burgers' equation, we were able to derive the average density of the flat-to-flat bridge explicitly
at certain specific values of intermediate times $t$, such
as $t=t_f/2$, $t=t_f/3$ and $t=t_f/4$. We also derive explicitly how the two edges of the
average density evolve from time $t=0$ to time $t=t_f$. Finally, we made links to other well studied problems, such as the Chern-Simons model, the related Stieltges-Wigert
orthogonal polynomials and the Borodin-Muttalib ensemble of determinantal point processes.   

We remark that there exists a well known simple case of the vicious Brownian bridges where both the initial ($t=0$) and the final ($t=t_f$) configurations
have densities given by the Wigner semi-circular law. In this ``Wigner-to-Wigner'' case, it is easy to see that the average density at all intermediate times 
remains a Wigner semi-circle (with a time dependent rescaling). This is because under the evolution via Dyson's Brownian motion $\beta = 2$ the average density
remains a semi-circle if the initial density is itself a Wigner semi-circle. In this paper, our result for the flat-to-flat geometry for the VBB provides another example
where one can make analytical progress. It would be of course very interesting to see if the VBB problem can be solved for other initial and final positions of the 
particles. This will be particularly useful to compute the large $N$ asymptotics of the Harish-Chandra-Itzykson-Zuber integral (for $\beta = 2$) connecting 
arbitrary matrices $A$ and $B$~\cite{BBMP2014:HCIZINTEGR,grela2020eikonal}.

Finally, our work derives the effective Langevin equation for vicious bridge configurations, thus generalising 
the single particle effective Langevin equation for a bridge \cite{MO2015:CONSTRAINTLANGEVIN} to an interacting many-body system. In fact, in the single
particle setting, an effective Langevin equation was derived not just for the bridge configuration but also for other
constrained walks such as the Brownian excursion, the Brownian meander, etc.  It would be interesting to extend the approach
presented here for the many-particle vicious bridges to that of vicious excursions or meanders.

\begin{acknowledgements}
We thank J.-P. Bouchaud, J. Bun, P. Mergny, H. Orland and M. Potters for useful discussions. This research was supported by ANR grant ANR-17-CE30-0027-01 RaMaTraF. 
JG acknowledges support by the TEAMNET POIR.04.04.00- 00-14DE/18-00 grant of the Foundation for Polish Science.
\end{acknowledgements}

\appendix

\section{Derivation of Eq. \eqref{chardeteq}}
\label{AppB}
We recall the definition of characteristic polynomial \eqref{chardet} and formulate its ``non-averaged'' version
\begin{align*}
\tilde{\Omega}_N(y) = \prod_{i=1}^N \left (e^y-x_i \right ).
\end{align*}
The positions $x_i$ evolve according to the Langevin equation \eqref{Langevinx} which is rewritten as the following stochastic differential equation (hereafter SDE)
\begin{align}
\label{sde}
dx_i = \frac{1}{N t_f} \sum_{j(\neq i)} \frac{x_i^2}{x_i - x_j} d \theta  + \frac{1+\theta}{t_f} x_i dW_i \;,
\end{align}
where the Wiener process is defined by $dW_i dW_j = \delta_{ij} \frac{1}{N} d\theta$. We compute the evolution equation for $\tilde{\Omega}_N(y)$ under the stochastic motion \eqref{sde}. To this end, we use Ito's lemma
\begin{align}
\label{sde2}
d\tilde{\Omega}_N = \sum_i \frac{\partial \tilde{\Omega}_N}{\partial x_i} dx_i + \frac{1}{2} \sum_{i,j} \frac{\partial^2 \tilde{\Omega}_N}{\partial x_i\partial x_j} dx_i dx_j \;.
\end{align}
We compute the derivatives as 
\begin{align*}
\frac{\partial \tilde{\Omega}_N}{\partial x_i} & = - \frac{\tilde{\Omega}_N}{e^y-x_i} \;, \\ 
\frac{\partial^2 \tilde{\Omega}_N}{\partial x_i \partial x_j} & = 0 \;, \\ 
\end{align*}
and find
\bea
&&\sum_i \frac{\partial\tilde{\Omega}_N}{\partial x_i} dx_i  = - \tilde{\Omega}_N \left ( \frac{N-1}{Nt_f} d\theta \sum_i \frac{x_i}{e^y-x_i} + \frac{1}{Nt_f} d\theta \sum_{j(\neq i)} \frac{x_i x_j}{(e^y-x_i)(x_i - x_j)} + \frac{1+\theta}{Nt_f} \sum_i \frac{x_i dW_i}{e^y-x_i} \right ), \label{plugback}\\
&&\sum_{i,j} \frac{\partial^2 \tilde \Omega_N}{\partial x_i\partial x_j} dx_i dx_j  = 0 \;. \nonumber
\eea
To obtain a closed equation, each term in the above expression ought to be expressed in terms of $\tilde{\Omega}_N$. To this end, we compute first partial results
\begin{align*}
\partial_y \tilde{\Omega}_N & = e^y \tilde{\Omega}_N \sum_i \frac{1}{e^y-x_i},\\ 
\partial_{yy} \tilde{\Omega}_N -\partial_y \tilde{\Omega}_N & = \tilde{\Omega}_N \sum_{i\neq j} \frac{e^{2y}}{(e^y-x_i)(e^y-x_j)}, 
\end{align*}
and use them to get
\bea
&&\sum_i \frac{x_i}{e^y-x_i} = -N + e^y \sum_i \frac{1}{e^y-x_i}  = - N + \frac{\partial_y \tilde{\Omega}_N}{\tilde{\Omega}_N}, \\
&&\sum_{i\neq j} \frac{x_i x_j}{(e^y-x_i)(x_i - x_j)} = \frac{N(N-1)}{2} + e^y(N-1) \sum_i \frac{1}{x_i - e^y} + \frac{e^{2y}}{2} \sum_{i\neq j} \frac{1}{(x_i - e^y)(x_j - e^y)} \nonumber \\
&& = \frac{N(N-1)}{2} - (N-1) \frac{\partial_y \tilde{\Omega}_N}{\tilde{\Omega}_N} + \frac{\partial_{yy} \tilde{\Omega}_N}{2\tilde{\Omega}_N} - \frac{\partial_{y} \tilde{\Omega}_N}{2\tilde{\Omega}_N} \;, 
\eea
which are plugged back into Eq. \eqref{plugback} and \eqref{sde2} to yield
\begin{align*}
d\tilde{\Omega}_N = & \frac{N-1}{2t_f}\tilde{\Omega}_N d\theta + \frac{1}{2Nt_f} \left ( \partial_y \tilde{\Omega}_N - \partial_{yy} \tilde{\Omega}_N \right ) d\theta - \frac{1+\theta}{Nt_f} \tilde{\Omega}_N  \sum_i \frac{x_i dW_i}{e^y-x_i}. 
\end{align*}
This is a closed equation for $\tilde \Omega_N$ and the stochastic part is proportional to $dW_i$. This thus drops out when looking at the averaged characteristic polynomial $\left < \tilde{\Omega}_N\right > = \Omega_N$, i.e.,
\begin{align*}
\partial_\theta \Omega_N = \frac{N-1}{2t_f}\Omega_N + \frac{1}{2Nt_f} \left ( \partial_y \Omega_N - \partial_{yy} \Omega_N \right ),
\end{align*}
which is exactly the equation \eqref{chardeteq} given in the text.

\section{Derivation of Eq. \eqref{rhoform}}
\label{appC}
We start from the Sochocki-Plemejl formula relating the particle density and the e-Green's function $H$
\begin{align}\label{start}
\rho^{(x)} (x,\theta) = -\frac{t_f}{\pi} \lim_{\epsilon \to 0_+} \text{Im} \left [  \frac{1}{w} \ln H(w;\theta) \right ]_{w = x-i\epsilon} \;.
\end{align}
We use the complex logarithm formula
\begin{align*} 
\ln(H(x-i\epsilon)) = \ln|H(x-i\epsilon)| + i \arg H(x-i\epsilon)) + 2k\pi i \;,
\end{align*}
where $k$ enumerates the branch cuts of the logarithm and where the real and imaginary parts are clearly separated. We also expand $1/(x-i\epsilon) = \frac{x+i\epsilon}{x^2+\epsilon^2}$, plug these two formulas into Eq. \eqref{start} and find
\begin{align*}
\rho^{(x)} (x,\theta) = -\frac{t_f}{\pi} \lim_{\epsilon \to 0_+} \left [ \frac{\epsilon \ln |H(x-i\epsilon)|}{x^2+\epsilon^2} + \frac{x}{x^2+\epsilon^2} ( \arg H(x-i\epsilon)+2k\pi) \right ] \;.
\end{align*}
We assume that $\frac{\epsilon \ln |H(x-i\epsilon)|}{x^2+\epsilon^2} \to 0$ as $\epsilon \to 0$ and we are left with 
\begin{align*}
\rho^{(x)} (x,\theta) = -\frac{t_f}{\pi} \lim_{\epsilon \to 0_+} \left [ \frac{x}{x^2+\epsilon^2} ( \arg H(x-i\epsilon)+2k\pi) \right ] \;.
\end{align*}
We choose the branch $k=0$ to make the resulting density normalizable (i.e. not divergent) and for $x>0$ the term $x/(x^2+\epsilon^2)$ is regular and can be safely taken out of the limit which finally gives Eq. \eqref{rhoform}.

\section{Calculation of the moments of the average density for $t=t_f/2$.}
\label{appD}

In this Appendix we compute the moments of the densities $\rho^{(x)}$ in Eq. (\ref{rhotheta1}) and $\rho^{(\lambda)}$ in Eq. (\ref{rholambda}). We start with $\rho^{(x)}$ and 
set $w_\pm^{(1)} = x_\pm$ in Eq. (\ref{rhotheta1}). The $k$-th moment of the average density, using Eq. (\ref{rhotheta1}), is then given by 
\begin{eqnarray}
m^{(x)}_{k,\theta=1} = \int_{x_-}^{x_+} dx\, x^k \rho^{(x)}(x;\theta=1) = \frac{t_f}{\pi} \int_{x_-}^{x_+} dx \, x^{k-1} \arctan \left [ \frac{\sqrt{4x-(1 + xT )^2}}{1+xT}\right ] \;.
\end{eqnarray}
We perform a change of variables $x = \frac{1}{T} \left ( \frac{2}{T} \left (1+ \sqrt{1-T}p \right ) - 1 \right )$, $dx = \frac{2\sqrt{1-T}}{T^2} dp$ resulting in 
\begin{align*}
m^{(x)}_{k,\theta=1} = \frac{2t_f \sqrt{1-T}}{\pi T^{2k}} \int_{-1}^1 dp \left ( 2p\sqrt{1-T} + 2-T \right )^{k-1} \arctan \left ( \frac{\sqrt{1-T}\sqrt{1-p^2}}{1+\sqrt{1-T}p} \right ).
\end{align*}
We continue with integration by parts. Since by normalization the zeroth moment is unity $m_{0,\theta=1}^{(x)} = 1$, we continue with $k \geq 1$ for which the following formula holds:
\begin{align*}
m^{(x)}_{k,\theta=1} = - \frac{t_f(2\sqrt{1-T})^{k-1}}{k\pi T^{2k}} \sum_{n=0}^{k-1} \binom{k-1}{n} \left ( \frac{2-T}{2\sqrt{1-T}} \right )^{k-1-n} \left [ (T-1) J_n - \sqrt{1-T} J_{n+1} \right ],
\end{align*}
where $J_n = \int_{-1}^1 dp p^n/\sqrt{1-p^2}$. These integrals are expressible via Catalan numbers $C_n = \frac{1}{n+1} \binom{2n}{n}$ since $J_0 = \pi$, $J_{2n} = \frac{2\pi(2n-1)}{4^n} C_{n-1}$ and  $J_{2n-1} = 0$ for $n>0$:
\begin{align}
m^{(x)}_{k,\theta=1} & = \frac{t_f (1-T)(2-T )^{k-1}}{kT^{2k}} + \frac{2t_f(1-T)^2(2-T)^{k-3}}{k T^{2k}} \sum_{n=0}^{\lfloor\frac{k-1}{2} \rfloor-1} \binom{k-1}{2n+2} \left ( \frac{\sqrt{1-T}}{2-T} \right )^{2n} (2n+1) C_{n} + \nonumber \\
& + \frac{t_f(1-T)(2-T)^{k-2}}{k T^{2k}} \sum_{n=0}^{\lfloor \frac{k}{2}-1 \rfloor} \binom{k-1}{2n+1} \left ( \frac{\sqrt{1-T}}{2-T} \right )^{2n}  (2n+1) C_{n} \;.  \label{catalanmoments}
\end{align}
Thus, interestingly, the Catalan numbers also appear here. 

An alternative formulation of these moments in terms of the modified Bessel function is possible with the use of an identity $\int_{-1}^1 dp \frac{e^{-\alpha p}}{\sqrt{1-p^2}} = \pi I_0(\alpha)$ applied to the integrals $J_n = (-1)^n \lim\limits_{\alpha \to 0} \frac{\partial^n}{\partial \alpha^n} \int_{-1}^1 dp \frac{e^{-\alpha p}}{\sqrt{1-p^2}}$. We find:
\begin{align*}
m_{k,\theta=1}^{(x)} & = \frac{t_f a (-2)^{k-1}}{k T^{2k}} \frac{\partial^{k-1}}{\partial \alpha^{k-1}} \left [ e^{-\frac{\alpha(2-T)}{2} } \left ( a I_0(\alpha a) - I_1 (\alpha a) \right ) \right ]_{\alpha=0}, \qquad k \geq 1,
\end{align*}
where $a = \sqrt{1-T}$ and $I_i(x)$ is the modified Bessel function. 

Using the relation \eqref{rel} between particle densities in $x$ and $\lambda$ spaces, we translate the moments find 
\begin{align*}
m^{(\lambda)}_{k,\theta=1} = \int_{\lambda_-}^{\lambda_+} d\lambda \lambda^k \rho^{(\lambda)}(\lambda;\theta=1) = \frac{t_f}{\pi} \left ( \frac{t_f}{2} \right )^k \int_{x_-}^{x_+} \frac{dx}{x} (\log x )^k \arctan \left ( \frac{\sqrt{4x-(1+xT)^2}}{1+xT} \right ).
\end{align*}
We again change the variables:
\begin{align*}
m^{(\lambda)}_{k,\theta=1} = \frac{t_f a}{2\pi(k+1)} \left ( \frac{t_f}{2} \right )^k \int_{-1}^1 dp  \frac{(c + \log (ap + b))^{k+1} (a+p) }{(ap+b)\sqrt{1-p^2}},
\end{align*}
with $a = \sqrt{1-T}, b = 1-T/2, c = \log 2 + 2/t_f$. Now with the identity $\int_{-1}^1 dp \frac{1}{(ap+b)^\alpha \sqrt{1-p^2}} = \frac{\pi}{\sqrt{b^2-a^2}^\alpha} P_\alpha \left ( \frac{b}{\sqrt{b^2-a^2}}\right )$ where $P_\alpha$ is a Legendre function, the moments in $\lambda$ space are given by:
\begin{align*}
m_{k,\theta=1}^{(\lambda)} = \frac{t_f(-1)^{k+1}}{2 (k+1)} \left ( \frac{t_f}{2} \right )^k \frac{\partial^{k+1}}{\partial \alpha^{k+1}}\left [ T^\alpha \left ( P_{\alpha-1} \left (\frac{2-T}{T} \right ) - P_{\alpha} \left (\frac{2-T}{T} \right ) \right ) \right ]_{\alpha = 0}. 
\end{align*}
As a last step, we propose a general identity (found only in special case $n=2$ in eq. 1.9 of \cite{Szm2017:LEGENDREFUNC} but checked by us numerically in Mathematica for $n$ up to 5):
\begin{align*}
\frac{\partial^n}{\partial \alpha^n} \left [ P_{\alpha-1} \left (x \right ) - P_{\alpha} \left (x \right ) \right ]_{\alpha = 0} = 
\begin{cases}
0, & n \text{ is even}\\
-2 \frac{\partial^n}{\partial \alpha^n} P_\alpha(x)_{\alpha=0}, & n \text{ is odd},
\end{cases}, 
\end{align*}
which renders the final expression:
\begin{eqnarray}
m_{k,\theta=1}^{(\lambda)} = \frac{t_f}{(k+1)} \left ( -\frac{t_f}{2} \right )^k \sum_{l=0}^{\left \lfloor \frac{k+1}{2} \right \rfloor } \binom{k+1}{l} (\log T)^{k-2l} \frac{\partial^{2l+1}}{\partial \alpha^{2l+1}}\left ( P_{\alpha} \left (\frac{2-T}{T} \right ) \right )_{\alpha = 0} \;. \label{moments_la}
\end{eqnarray}
Although the expression for the moments in the $x$-space in (\ref{catalanmoments}) is easy to evaluate numerically, evaluating explicitly the moments in the $\lambda$-space from Eq. (\ref{moments_la}) is difficult due to the apparent lack of explicit formulae for the derivatives of the Legendre function with respect to its degree (see e.g. \cite{Szm2017:LEGENDREFUNC}).

\section{Computation of the support for the $\beta=2$ Dyson's Brownian Motion}\label{App:E}

We consider the Dyson's Brownian motion with $\beta=2$ where the positions $\lambda_i(t)$ of $N$ particles evolve in time via Eq. (\ref{model}), starting from the initial positions $\vec{a}$ at $t=0$. One defines the Green's function $G_N(z,t)$
\bea \label{def_Gzt}
G_N(z;t) = \frac{1}{N} \sum_{i=1}^N \frac{1}{z-\lambda_i(t)} \;.
\eea
The average density can be obtained via the Sochocki-Plemelj formula
\bea \label{SP_formula}
\rho(\lambda;t) = \frac{1}{\pi} \,{\rm Im}\,G_N(z - i 0^+;t) \;.
\eea
In the large $N$ limit, this converges to
\bea \label{resolvent}
G(z;t) = G_\infty(z,t) = \int \frac{\rho(\lambda;t)}{z-\lambda} \, d\lambda \;,
\eea
where $\rho(\lambda;t)$ is the average density at time $t$. Thus the initial condition for $G(z;t)$ reads 
\bea \label{IC_G}
G(z;t=0) = G_0(z) =  \int \frac{\rho(\lambda;0)}{z-\lambda} \, d\lambda \;,
\eea
where $\rho(\lambda;0)$ is the initial density. In the large $N$ limit, one can show that $G(z,t)$ satisfies the inviscid Burgers' equation \cite{BN2010:SHOCKRMT,allez2012invariant,BGNW2015:HERMDETS,krajenbrink2020tilted}
\bea \label{Burgers_GUE} 
\partial_t G(t) + G(z,t) \partial_z G(z,t) = 0 \;.
\eea
The solution can be obtained in the parametric form by the method of characteristics, as discussed in Section \ref{sec2}. It reads
\bea \label{param_GUE}
G(z;t) = G_0(\xi) \;,
\eea
where $\xi$ and $z$ are related by
\bea \label{z_vs_xi_GUE}
z = \xi + t \,G_0(\xi) \;.
\eea
Given $z$ and $t$, we need to solve Eq. (\ref{z_vs_xi_GUE}) for $\xi$ and then substitute in Eq. (\ref{param_GUE})
to obtain $G(z;t)$. 

\begin{figure}[t]
\includegraphics[width = 0.5\linewidth]{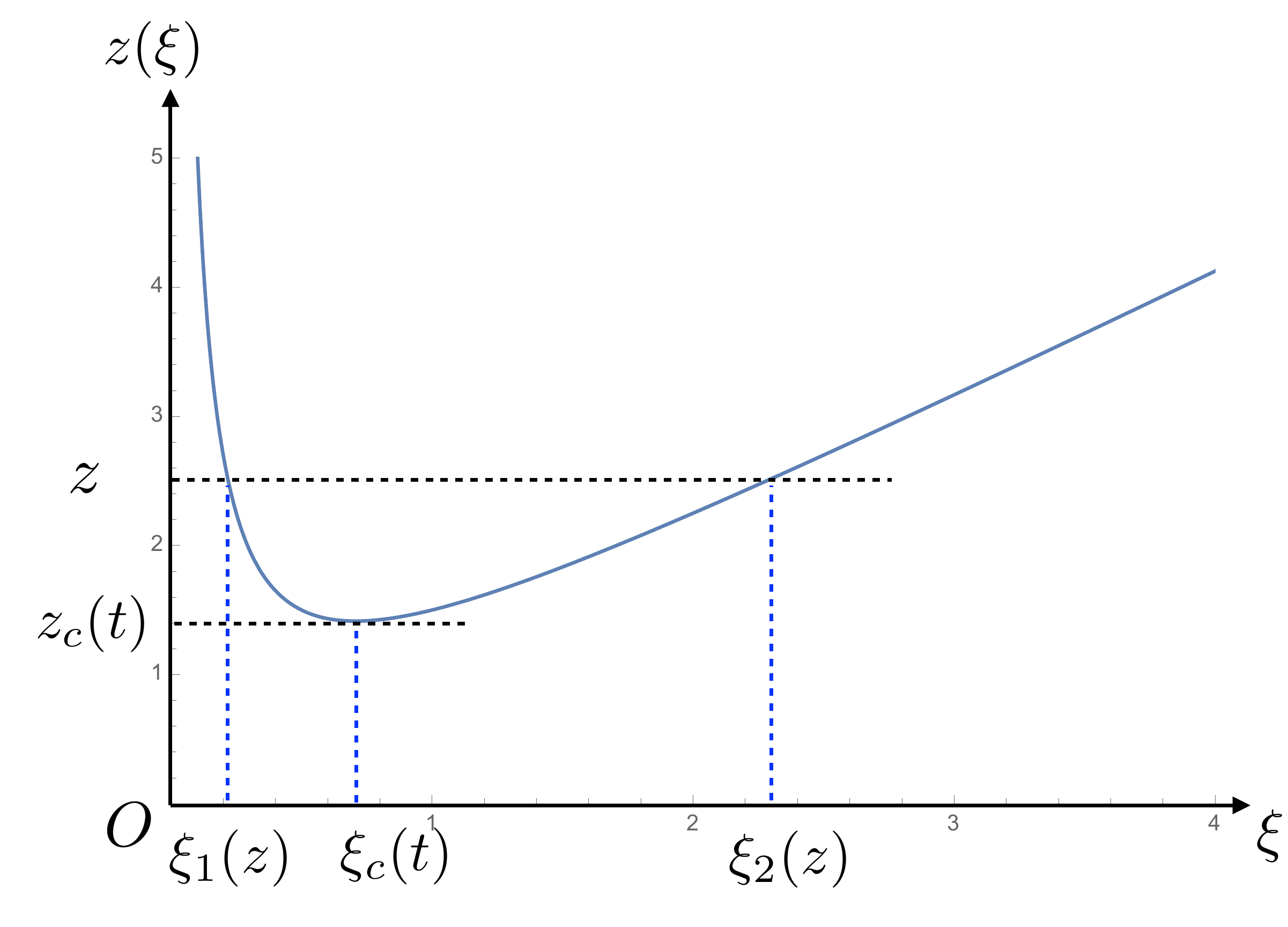}
\caption{Plot of $z(\xi)$ in Eq. (\ref{z_vs_xi_GUE}) for $G_0(\xi) = 1/\xi$ and $t=1/2$.} \label{Fig_Burgers}
\end{figure}

Let us first discuss some general properties of the Green's function $G(z;t)$. 
Consider first $G(z;t)$ as a function of $z$ along the positive real
axis. An exactly similar analysis can be done on the negative real axis. 
From the definition in Eq. (\ref{def_Gzt}), it is clear that as $z \to \infty$, 
$G(z;t) \simeq 1/z$ since $\rho(\lambda;t)$ is normalized to $1$. As $z$ decreases from $+\infty$,
$G(z;t)$ typically increases with decreasing $z$. However, this does not go on for ever since we expect that
there will be a cut along the real axis where $G(z;t)$ acquires a nonzero imaginary part, which gives rise to
a nonzero density via Eq. (\ref{SP_formula}). Let $z_{\rm edge}(t)$ denote this value of $z$ below which $G(z;t)$
is imaginary. Clearly, this is also the {\it upper edge} of the support of the density. Thus in the range $[z_{\rm edge}(t), +\infty)$, $G(z;t)$
is a monotonically decreasing function of $z$. Similarly, the function $G_0(\xi)$ is, generically, a monotonically decreasing
function of $\xi$ for $\xi \in [\xi_{\rm edge}, +\infty)$ and decays for large $\xi$ as $G_0(\xi) \simeq 1/\xi$. Note that, for simplicity, we have
presented only the behavior for the upper edge of the support of the density. A similar analysis can be done for the lower edge of the support,
for which we need to analyse the Green's function $G(z;t)$ on the negative real axis in the complex $z$-plane.

Given $z$ and $t$, we now want to find the solution for $\xi$ from Eq. (\ref{z_vs_xi_GUE}). Suppose that we plot $z$ as a function
of $\xi$ for $\xi > \xi_{\rm edge}$. The rhs of (\ref{z_vs_xi_GUE}) is the sum of two terms: the first one increases linearly with $\xi$ while
the second term, for fixed $t$, is a decreasing function of $\xi$. Hence their sum, when plotted as a function of $\xi$ will first decrease with
increasing $\xi$, achieves a minimum at $\xi_c(t)$ and then increases monotonically with $\xi$ (see Fig. \ref{Fig_Burgers}). For a given $z$ and $t$, 
this equation (\ref{z_vs_xi_GUE}) has typically two solutions $\xi_1(z) < \xi_2(z)$. Here, $\xi_2(z)$ is a monotonically increasing function of $z$, while
$\xi_1(z)$ is a monotonically decreasing function of $z$ (see Fig. \ref{Fig_Burgers}). The correct solution is actually given by $\xi_2(z)$. This 
is because from Eq. (\ref{param_GUE}) we see that since $G(z;t)$ is a monotonically decreasing function of $z$ and $G_0(\xi)$ is also
a monotonically decreasing function of $\xi$, hence $z$ must be a monotonically increasing function of $\xi$. This justifies the fact that
$\xi_2(z)$ is the correct root. Note however that $\xi_2(z)$ exists only if $z \geq z_c(t)$ where $z_c(t)$ is the value of
$z(\xi) = \xi + t \, G_0(\xi)$ at the minimum located at $\xi = \xi_c(t)$ (see Fig. \ref{Fig_Burgers}). This minimum is obtained by setting
\bea \label{dz}
\frac{dz}{d\xi} = 1 + t \, G_0'(\xi) = 0  \;.
\eea  
This gives the value $\xi_c(t)$ and consequently 
\bea \label{zc}
z_c(t) = \xi_c(t) + t \, G_0(\xi_c(t)) \;.
\eea  
Therefore we see that the real solution exists only for $z \geq z_c(t)$. Thus we identify 
\bea \label{ze_zc}
z_{\rm edge}(t) = z_c(t) \;.
\eea

As an example, let us consider the simple case where all the particles start at the origin, i.e. $\rho(\lambda;0) = \delta(\lambda)$. In this case,
from Eq. (\ref{IC_G}) we have $G_0(\xi) = 1/\xi$. In this case $\xi_{\rm edge} = 0$. Consequently, from Eq. (\ref{dz}), we get
\bea \label{dz_GUE}
\xi_c(t) = \sqrt{t} \;.
\eea
Note that $\xi_c(t)$ has nothing to do with $\xi_{\rm edge}$. Here we are considering the positive side of the support, hence we choose the
positive root in (\ref{dz_GUE}). Similarly, when analyzing the lower edge of the support, we should instead choose the negative root 
$\xi_c(t) = - \sqrt{t}$. From Eq. (\ref{zc}) one gets for the positive side 
\bea \label{zc_GUE}
z_c(t) = \sqrt{4\,t} \;.
\eea 
This result tells us how the upper support of the average density evolves with time $t$. Indeed, in this case, 
one can solve for $\xi$ from the quadratic equation $z = \xi  + t/\xi$, which gives two roots: $\xi_1(z) = (z - \sqrt{z^2-4})/2$
and $\xi_2(z) = (z + \sqrt{z^2-4t})/2$. As argued earlier, we choose the second branch as the correct one, i.e., $\xi = (z + \sqrt{z^2-4t})/2$.
Consequently, from Eq. (\ref{param_GUE}) one gets
\bea \label{Gzt_GUE}
G(z;t) = G_0(\xi) = \frac{2}{z + \sqrt{z^2-4t}} = \frac{1}{2t} \left(z - \sqrt{z^2-4t} \right) \;.
\eea
Therefore, from the relation (\ref{SP_formula}), the average density at time $t$ is given by the semi-circular form
\bea \label{Wigner}
\rho(\lambda;t) = \frac{1}{2\pi t} \sqrt{4t - \lambda^2} \;.
\eea 
Hence we see that the density is supported over the interval $[-\sqrt{4t},+\sqrt{4t}]$ and indeed the upper support $\sqrt{4t}$ coincides with 
the value of $z_c(t)=z_{\rm edge}(t)$ in Eqs.  (\ref{ze_zc}) and (\ref{zc_GUE}).

\bibliographystyle{apsrev4-1}
\bibliography{krkbib2015}{}

\end{document}